\documentclass[longauth]{aa}
\usepackage{txfonts}
\usepackage{graphicx}
\usepackage{natbib}
\usepackage{color}
\bibpunct{(}{)}{;}{a}{}{,} 

\begin{document}

\renewcommand{\cite}{\citep} 


\newcommand{\HMS}[3]{$#1^{\mathrm{h}}#2^{\mathrm{m}}#3^{\mathrm{s}}$}
\newcommand{\DMS}[3]{$#1^\circ #2' #3''$}
\newcommand{\StatSysErr}[3]{#1 \pm #2_{\mathrm{stat}} \pm #3_{\mathrm{sys}}}

\newcommand{\TODO}[1]{\textbf{\texttt{\textcolor{green}{\emph{[#1]}}}}}
\newcommand{\UPDATED}[1]{\textbf{\emph{#1}}}
\newcommand{\REMOVED}[1]{\textcolor{red}{[OMIT #1]}}
\newcommand{\Jim}[1]{\textcolor{red}{\textbf{\emph{[JIM$>$ #1]}}}} 
\newcommand{\UNITS}[1]{\,\mathrm{#1}}

\newcommand{\CO}{$^{12}$CO}
\newcommand{\KMPERSEC}{\mathrm{km\,s^{-1}}}
\newcommand{\XMM}{\emph{XMM-Newton}}
\newcommand{\HESS}{H.E.S.S.}

\newcommand{\HESSINDEX}{\StatSysErr{2.71}{0.11}{0.2}} 
\newcommand{\HESSNORM}{(\StatSysErr{2.84}{0.23}{0.28})\times10^{-12}\UNITS{TeV^{-1}\,cm^{-2}\,s^{-1}} } 
\newcommand{\HESSINTFLUX}{F(1-10\UNITS{TeV})=(1.63 \pm 0.16)\times10^{-12}\UNITS{cm^{-2}\,s^{-1}}} 

\newcommand{\HESSJ}{HESS~J1745$-$303}
\newcommand{\SNRa}{G359.1$-$0.5}
\newcommand{\SNRb}{G359.0$-$0.9}
\newcommand{\HESSGC}{HESS~J1745$-$290}
\newcommand{\EGRSRC}{3EG~J1744$-$3011}
\newcommand{\PSRa}{PSR~B1742$-$30}
\newcommand{\PSRb}{PSR~J1747$-$2958}


\title{Exploring a SNR/Molecular Cloud Association Within \HESSJ}
\titlerunning{\HESSJ\ }
\authorrunning{The \HESS\ Collaboration}
\author{F. Aharonian\inst{1,13}
 \and A.G.~Akhperjanian \inst{2}
 \and U.~Barres de Almeida \inst{8} \thanks{supported by CAPES Foundation, Ministry of Education of Brazil}
 \and A.R.~Bazer-Bachi \inst{3}
 \and B.~Behera \inst{14}
 \and M.~Beilicke \inst{4}
 \and W.~Benbow \inst{1}
 \and K.~Bernl\"ohr \inst{1,5}
 \and C.~Boisson \inst{6}
 \and O.~Bolz \inst{1}
% ...
}

\institute{
Max-Planck-Institut f\"ur Kernphysik, P.O. Box 103980, D 69029
Heidelberg, Germany
\and
 Yerevan Physics Institute, 2 Alikhanian Brothers St., 375036 Yerevan,
Armenia
\and
Centre d'Etude Spatiale des Rayonnements, CNRS/UPS, 9 av. du Colonel Roche, BP
4346, F-31029 Toulouse Cedex 4, France
\and
Universit\"at Hamburg, Institut f\"ur Experimentalphysik, Luruper Chaussee
149, D 22761 Hamburg, Germany
\and
Institut f\"ur Physik, Humboldt-Universit\"at zu Berlin, Newtonstr. 15,
D 12489 Berlin, Germany
\and
LUTH, Observatoire de Paris, CNRS, Universit\'e Paris Diderot, 5 Place Jules Janssen, 92190 Meudon, 
France
\and
DAPNIA/DSM/CEA, CE Saclay, F-91191
Gif-sur-Yvette, Cedex, France
\and
University of Durham, Department of Physics, South Road, Durham DH1 3LE,
U.K.
\and
Unit for Space Physics, North-West University, Potchefstroom 2520,
    South Africa
\and
Laboratoire Leprince-Ringuet, Ecole Polytechnique, CNRS/IN2P3,
 F-91128 Palaiseau, France
\and 
Laboratoire d'Annecy-le-Vieux de Physique des Particules, CNRS/IN2P3,
9 Chemin de Bellevue - BP 110 F-74941 Annecy-le-Vieux Cedex, France
\and
Astroparticule et Cosmologie (APC), CNRS, Universite Paris 7 Denis Diderot,
10, rue Alice Domon et Leonie Duquet, F-75205 Paris Cedex 13, France
\thanks{UMR 7164 (CNRS, Universit\'e Paris VII, CEA, Observatoire de Paris)}
\and
Dublin Institute for Advanced Studies, 5 Merrion Square, Dublin 2,
Ireland
\and
Landessternwarte, Universit\"at Heidelberg, K\"onigstuhl, D 69117 Heidelberg, Germany
% ...
}

\offprints{K. Kosack, \email{Karl.Kosack@mpi-hd.mpg.de}, A. Fiasson
  \email{Armand.Fiasson@LPTA.in2p3.fr}}

\abstract{}{\HESSJ\ is an extended, unidentified VHE (very high
  energy) gamma-ray source discovered using \HESS\ in the Galactic
  Plane Survey. Since no obvious counterpart has previously been found
  in longer-wavelength data, the processes that power the VHE
  emission are not well understood.}{Combining the latest VHE data
  with recent \XMM\ observations and a variety of source catalogs and
  lower-energy survey data, we attempt to match (from an energetic and
  positional standpoint) the various parts of the emission of \HESSJ\
  with possible candidates.}{Though no single counterpart is found
  to fully explain the VHE emission, we postulate that at least
  a fraction of the VHE source may be explained by a
  supernova-remnant/molecular-cloud association and/or a
  high-spin-down-flux pulsar.}{}


\maketitle


\section{Introduction}

The Galactic center region (roughly between $|l|<2^\circ$,
$|b|<1^\circ$) is densely populated with possible VHE emission
candidates: supernova remnants (SNRs), dense molecular clouds, pulsar
wind nebulae (PWNe), X-ray binaries (XRBs), and a variety of
unidentified sources seen in lower wavebands. In this region, VHE
emission has been detected from an as yet unidentified point-like
source at the Galactic center (possibly associated with the
super-massive black hole Sgr~A$^{\star}$ or a PWN)
\cite{tsuchiya04:_detec_sub_tev_gamma_rays,kosack04,HESS:GC,hinton07},
from the SNR G\,0.9$+$0.1 \cite{HESS:G09} (point-like for H.E.S.S.),
and from a region of diffuse emission approximately $\pm 1^\circ$ in
longitude, which is most likely associated with the interaction of
cosmic-ray particles with molecular clouds \cite{HESS:gc_diffuse}, and
finally \HESSJ, an extended, unidentified VHE gamma-ray source lying
approximately a half degree below the Galactic plane at $l=-0.4$.
\HESSJ\ was first discovered in the \HESS\ Galactic Plane Survey
\cite{HESS:scanpaper2}. Subsequent observations of the region using
\HESS\ have provided increased exposure of this object, and thus a
more detailed study is now possible.

VHE gamma rays are typically thought to be produced via two general
mechanisms: the up-scatter of lower-energy photons by high-energy
electrons via the inverse-Compton process, or the production and
subsequent decay of $\pi^0$s produced in the interactions of
high-energy hadrons. Though it is difficult to distinguish between a
purely leptonic or hadronic scenario in many of the currently
published VHE sources, in cases where it is known that cosmic rays are
interacting with a dense medium, the hadronic scenario becomes more
viable. This is true for example on the Galactic center ridge, where
the VHE emission is seen to roughly follow the location of dense
molecular clouds \cite{HESS:gc_diffuse}, or possibly in the case of
SNRs embedded in dense regions of the interstellar medium (ISM)
\cite[e.g.][]{HESS:W28,HESS:RXJ1713}.  The flux of gamma rays produced
via the hadronic production of $\pi^0$s depends linearly on the
density of the surrounding medium. For typical Galactic SNRs (few kpc
distance, age of $> 1000\,$yrs, in a medium with average density
$\sim1\UNITS{cm^{-3}}$), the flux above 1~GeV is predicted to be quite
small (below the EGRET or even \HESS\ sensitivity)
\cite{aharonian94:_gev_tev_gamma_ray_emiss}. However, in the case
where the supernova shock is interacting with a dense molecular cloud
(as in the case of \SNRa, discussed later), the emission can be
significantly enhanced. Moreover, the expected flux in the TeV energy
range from an interacting SNR shock may be much higher than the
extrapolation of the spectrum measured in the GeV range
\cite{aharonian96:_emiss_of_pi_decay_gamma}.

Though the region around \HESSJ\ is well covered by radio
\cite[e.g. VLA,][]{larosa00:_wide_field_centim_vla_image}, and X-ray
\cite[e.g. ROSAT,][]{ROSAT} observations, no obvious counterpart is visible
that fully matches the morphology of this source. Here, we examine the
possibility that \HESSJ\ is (at least in part) associated with several
counterpart candidates seen in other wavebands, most notably the
interaction of a nearby supernova remnant with a molecular cloud.
Additionally, we present an analysis of recent X-ray data from
\XMM\ covering the central part of this object.

\section{Technique}

\subsection{The \HESS\ Instrument}

\HESS\ (the High Energy Stereoscopic System) is an array of four
atmospheric Cherenkov telescopes (ACTs) located in the Khomas
highlands of Namibia at an altitude of $1800\UNITS{m}$ above
sea-level. Each telescope consists of a $107\UNITS{m^2}$ optical
reflector made up of segmented spherical mirrors that focus light into
a camera of 960 photo-multiplier tube pixels \cite{HESS:optics}. Using
the \emph{imaging atmospheric Cherenkov technique}
\citep[e.g.][]{hillas85,hillas96:technique,weekes96:acts,HEGRA:acts},
the telescopes image the Cherenkov light emitted by the particles in
extensive air showers from multiple viewpoints, and the energy and
direction of the primary gamma ray can be reconstructed with an
average energy resolution of $\sim$16\% (above an energy threshold of
approximately 150 GeV), and a spatial resolution of $\sim{\hskip
-0.3em}0.1^\circ$ per event \cite{HESS:crab}. The large field of view
($\sim{\hskip -0.3em}5^\circ$), and good off-axis sensitivity of the
\HESS\ array make it well suited for studying extended sources and for
scan-based observations, where the source position is not known a
priori.

\subsection{Data and Analysis Technique} 
\label{sec:technique}

Following the standard \HESS\ procedure, the data presented here are
processed with separate analysis and calibration schemes: the \HESS\
\emph{standard analysis} \cite{HESS:crab}, in which showers are
reconstructed and hadronic background is rejected via the Hillas
moment-analysis technique \cite{hillas96:technique}; and the
\emph{Model2D} analysis described by \citet{HESS:Model2D}, which
employs a semi-analytic model of the shower to characterize each
image. It should be noted that both the techniques and simulations
used in them are independent, providing a robust check of the
analysis. Since results of both analyses agree within errors, the
results presented here are from the \emph{standard analysis} only. The
separation of gamma-ray candidates from cosmic-ray-like events was
made using both \emph{standard cuts} (optimized for a lower energy
threshold) and \emph{hard cuts} (optimized for better background rejection)
described in \cite{HESS:crab}, but with a larger angular angular
integration radius to account for the extension of the source. The
former were used for spectra, and the latter for producing the sky
images, though both were checked for consistency.

The data for \HESSJ\ are not primarily comprised of dedicated
observations of the object, but rather from scan-based observations of
the region (taken at regular grid-points along the galactic plane) and
from dedicated observations of the VHE source at the Galactic center
(\HESSGC), which lies approximately $1.4^\circ$ away. Due to the wide
range of pointings and since the gamma-ray acceptance across the field
of view of \HESS\ falls off radially, the total
exposure of the region around \HESSJ\ is highly non-uniform; in
particular there is a strong gradient toward the Galactic center,
which can lead to increased systematic errors in background
subtraction. To mitigate this situation, several different background
selection techniques were employed in this analysis.

For the generation of two-dimensional images, both the
\emph{field-of-view background} method, where the background for each
observation is determined from a one-dimensional model of the radial
acceptance (taken from observations with no significant emission in
the field of view), and the \emph{ring background} method, where the
background at each point on the sky is calculated from an annulus
surrounding it (again, with sources excluded), are used
\cite{HESS:background}.  Since the \emph{field-of-view background}
method is more sensitive to gradients in exposure, the images
presented here employ the \emph{ring background} method, though both
methods are checked for reasonable (within $\sim$5\%) consistency.

For the spectral analysis background estimation, we use the
\emph{reflected region} technique \cite{HESS:background}, where
background events are selected from circular off-source regions within the
field of view. These regions are chosen with the same angular size and
offset from the observation center position as the on-source
integration region, ensuring their acceptance-corrected exposure is
approximately equal to that of the on-source region. This technique is
less suited to the generation of images, but for a known source
position provides an estimate of the background that is 
independent of radial acceptance models.

To prevent contamination of the background from the diffuse gamma-ray
emission and to avoid including emission from the various other
sources near the Galactic center, regions around Sagittarius~A$^{\star}$,
around G\,0.9$+$0.1, as well as within $\pm 0.8^\circ$ of the
Galactic plane, were not used for background estimation.

The statistical significances in both the images and the spectral
analysis are calculated from the measured number of on- and off-source
(background) events following the likelihood ratio procedure outlined
in \citet{li_ma83}.

Spectra are generated following the methods described by both
\citet{HESS:crab}, in the case of the \emph{standard analysis}, and
\citet{piron01:CAT_ForwardFolding_Mrk421} for the \emph{Model2D
  analysis}. The on-source integration radius used for the generation
of spectra is chosen (unless otherwise stated) to fully enclose the
source (based on the radial profile of the signal and background), 
making no assumptions on the details of the source morphology
and thus providing a less-biased result at the expense of a lower
signal-to-noise ratio. Since the data set spans several years, during
which the gain and optical efficiency of the telescopes has changed,
muon images are used to calibrate the energy estimate of each
gamma-ray candidate \cite{HESS:crab}. The systematic error on the
fluxes given here is estimated from simulated data to be $\pm$20\%
while the photon index has a typical systematic error of $\pm0.2$.

\section{VHE Results}
\label{sec:results}

 \begin{figure}
   \centering
   \resizebox{0.9\hsize}{!}{\includegraphics{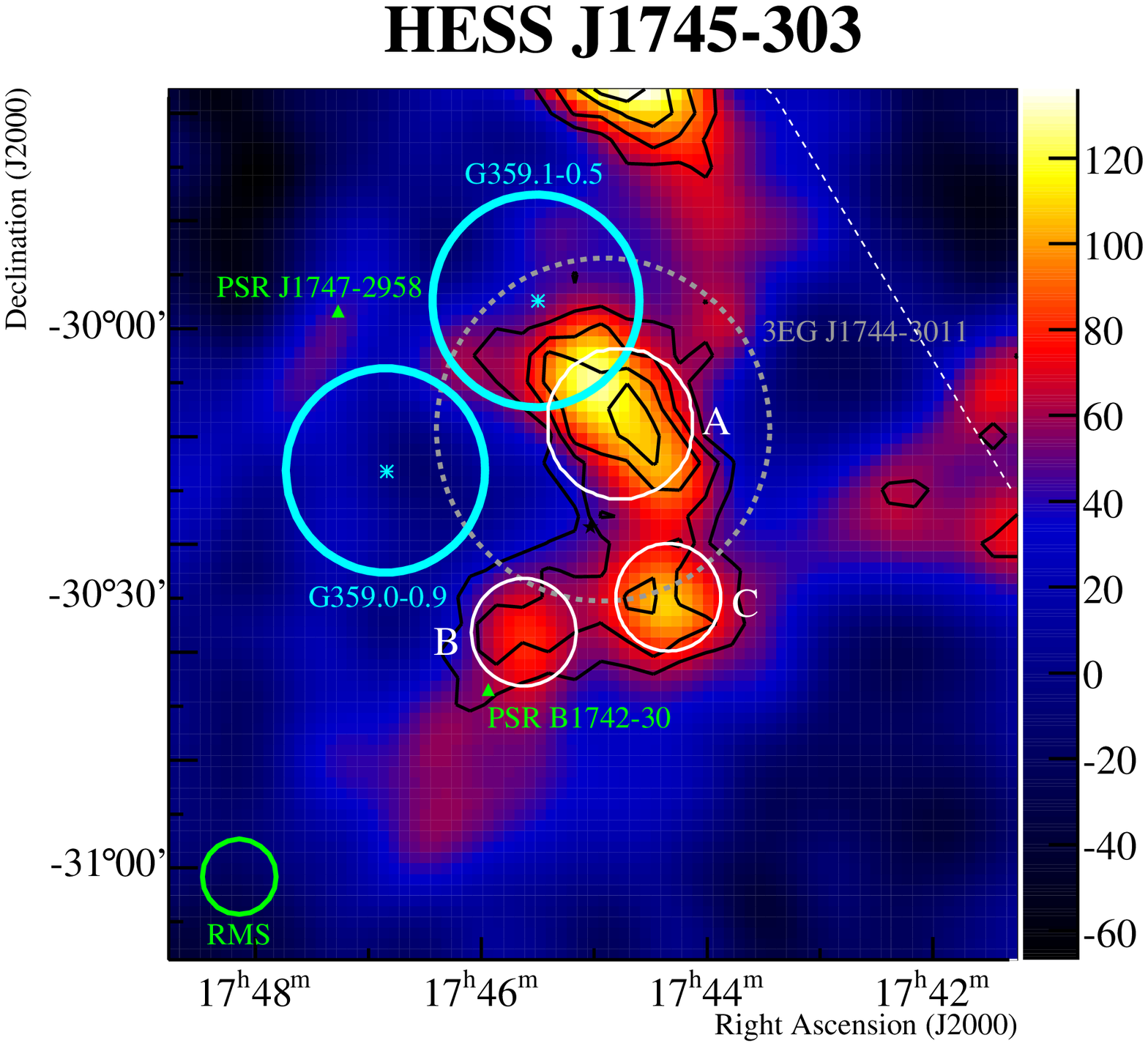}}
   \resizebox{0.9\hsize}{!}{\includegraphics{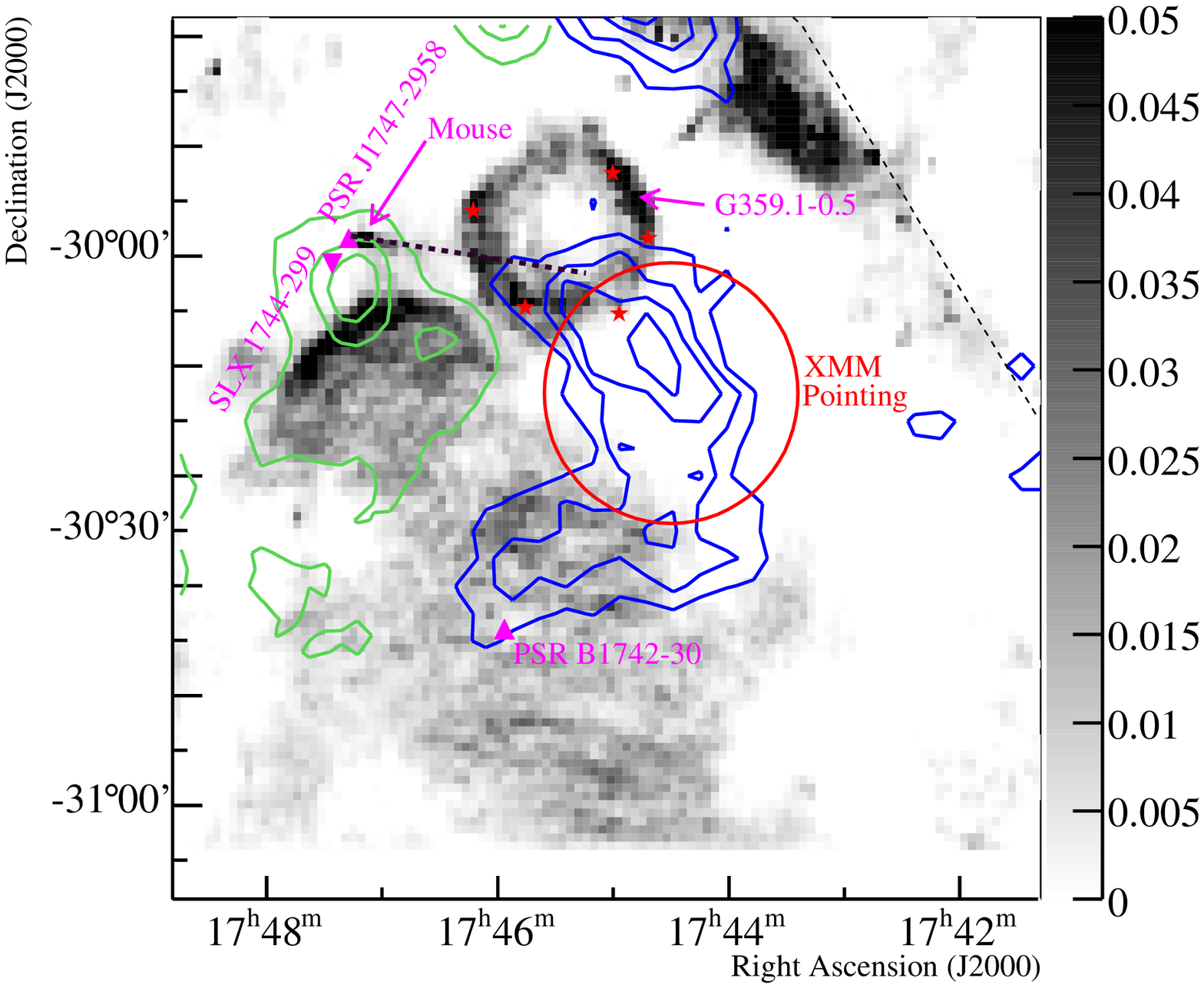}}
   \caption{ \emph{Top:} A VHE gamma-ray (excess count) image of
     HESS~J1745-303 with the positions of possible counterpart
     candidates overlaid for reference.  The color scale set such that
     the blue/red transition occurs at the $\sim3\sigma$ (pre-trials)
     significance level. The $4\sigma$ to $7\sigma$ statistical
     significance contours are shown in black. The thin white circles
     represent the integration regions \emph{A}, \emph{B}, and
     \emph{C} discussed in the text. The dashed circle is the 95\%
     error circle for the location of \EGRSRC. The significant excess
     seen to the north is the tail end of the Galactic Ridge diffuse
     emission discussed by \citet{HESS:gc_diffuse}. The Galactic plane
     is marked with a dotted line. \emph{Bottom:} The \HESS\
     significance contours (blue) overlaid on a VLA radio image
     \cite{larosa00:_wide_field_centim_vla_image} with overlaid ROSAT
     hard-band contours \cite[green,][]{ROSAT}. Stars show the
     positions of OH masers, and the \XMM\ field of view (Figure
     \ref{fig:xmmfov}) is drawn as a circle.}
   \label{fig:VHE}
 \end{figure}

 \begin{figure}
   \centering
   \resizebox{\hsize}{!}{\includegraphics{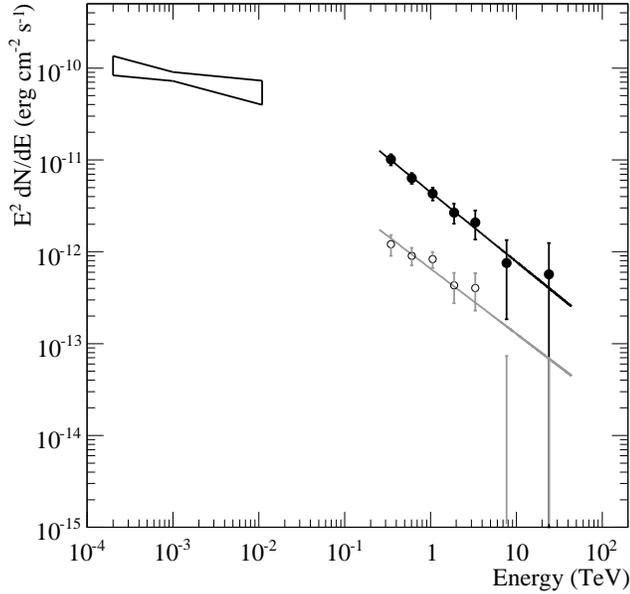}}
   \caption{ The spectral energy distribution of VHE gamma rays from
     \HESSJ\ (solid points) and from only region \emph{A} (open
     circles). The data are fit by a power law of the form $dN/dE =
     N_0 (E/TeV)^{-\Gamma}$ see Section \ref{sec:results}. The \EGRSRC\
     flux is plotted as a bow-tie for comparison.  }
   \label{fig:spect}
 \end{figure}

 With an exposure time of 35 hours, the original \HESS\ data set
 presented by \citet{HESS:scanpaper2} revealed \HESSJ\ at a pre-trials
 significance level of 6.3 standard deviations (for an integration
 radius of $0.224^{\circ}$, which is the \HESS\ standard for blind
 source searches). With the increased exposure (now 79 hours) coming
 primarily from re-observations of the Galactic center source \HESSGC,
 \HESSJ\ is now seen at well above the detection threshold, with a
 pre-trials significance level of 12 standard deviations ($\sigma$).

 Figure \ref{fig:VHE} shows an image of gamma-ray excess counts
 covering \HESSJ\ with significance contours overlaid. The image is
 smoothed with a Gaussian kernel with standard deviation $0.07^\circ$
 (chosen to reveal morphological features while maintaining good
 statistics), and the significance contours are generated with an
 oversampling radius of $0.12^\circ$, matched to the RMS of the
 Gaussian smoothing to provide a visual impression of significant
 features. The source centroid is determined by fitting an elongated
 two-dimensional Gaussian convolved with the \HESS\ point-spread
 function to the un-smoothed images. Due to the non-Gaussian 
 morphology, this only gives a rough centroid of the emission.

 \begin{figure}
   \centering
   \resizebox{\hsize}{!}{\includegraphics{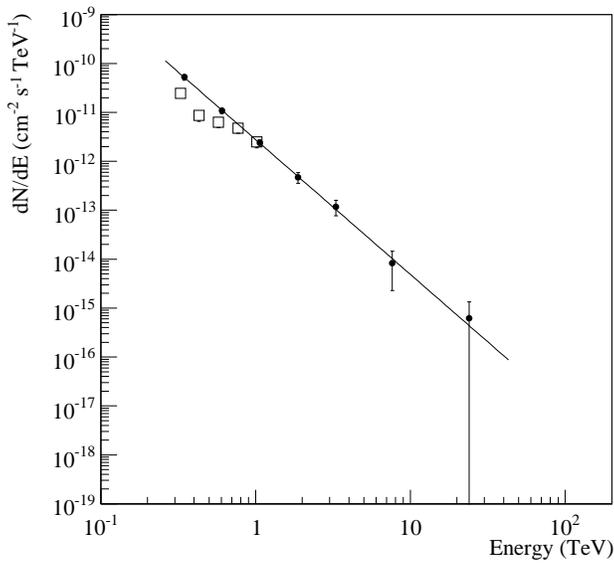}}
   \label{fig:sp2compare}
   \caption{Comparison of the spectra of \HESSJ\ presented here
     (filled circles with line fit) with the previous result presented
     by \citet{HESS:scanpaper2} (open squares).  }
 \end{figure}
 
 For the spectral analysis, the integration region was taken to
 include the entire source (with a radius of $0.4^\circ$ centered on
 \HMS{17}{45}{2.10}, \DMS{-30}{22}{14.00}, J2000 coordinates). For
 this larger integration region, we find a total significance of
 10.2$\sigma$, with 2030 total excess counts. The spectrum, shown in
 Figure \ref{fig:spect}, is well fit by a power law: $dN/dE = N_0
 (E/\mathrm{TeV})^{-\Gamma}$ with photon index $\Gamma=\HESSINDEX$,
 and a differential flux normalization $N_0$ of $\HESSNORM$. This
 corresponds to an integral flux above the peak energy of VHE events
 ($\sim$1 TeV) of $\HESSINTFLUX$. Figure \ref{fig:spect} shows the
 resulting spectral energy distribution.

 \begin{table}
   \begin{center}
     \begin{tabular}{r c c c}
       \hline\hline
       & \HESS & \HESS & \XMM \\
       & (2006) & (full) & \\
       \hline
       Exposure Time (h) & 35 & 79 & 4 \\
       Angular Res. (arcmin) & 6 & 6 & 0.1\\
       Field-of-view ($^\circ$) & 5.0 & 5.0 &  0.5 \\ 
       Energy Range & 0.35-1.5~TeV   &  0.35-30~TeV & 0.5-10 keV\\
       Energy Res. & 16\% & 16\% & 20-50\%\\
       \hline\hline
     \end{tabular}
   \end{center}
   \caption{Characteristics of the observations of \HESSJ\ with \HESS\
     \cite{HESS:crab} and \XMM. The ``full'' dataset is what is
     presented here, while the 2006 subset was presented by
     \citet{HESS:scanpaper2}.  The energy range for the \HESS\ case is
     the range used to fit the statistically significant spectral
     points; the full sensitivity of the detector extends to
     approximately 100 TeV. }
 \end{table}




 Figure \ref{fig:sp2compare} compares the new result to the data
 presented in \citet{HESS:scanpaper2}. Below 700 GeV, the new result
 noticeably differs from the former spectrum, which was made with
 nearly a factor of 10 less photons (211 excess
 counts). Quantitatively, the new spectrum is 2.5$\sigma$ softer than
 the old photon index of $\Gamma=1.82 \pm 0.29_\mathrm{stat} \pm
 0.2_\mathrm{sys}$. Thus, the integral flux above 200 GeV is smaller
 than the old value of $(\StatSysErr{11.2}{4.0}{3.4}) \times
 10^{-12}\UNITS{cm^{-2}s^{-1}}$.

 A re-analysis of only the data used in the original publication with
 our updated procedures gives a result that is in agreement with the
 one presented here; and a re-analysis of all other sources presented
 by \citet{HESS:scanpaper2} using the same techniques as in this paper
 gives results that are consistent with the original publication.
 Therefore, we have no indication that there is a general systematic
 error involved in either the old or the new analysis, but rather most
 likely a systematic error that exclusively (or predominantly)
 affects \HESSJ.

 When analyzing an extended source that has both
 low-surface-brightness and is located in a region of extremely uneven
 exposure, small uncertainties, e.g. in the acceptance-correction or
 background subtraction, may become significant. Such a problem only
 strongly affects low-surface-brightness sources analyzed with large
 integration radii. For most \HESS\ sources, wobble-mode observations
 are preformed that deliberately constrain systematic background
 errors to a minimum; but in the case of HESS J1745-303, the data set
 is dominated by observations of the Galactic center and the Galactic
 plane, therefore a uniform acceptance for all background control
 regions and the source region was probably not achieved in the old
 analysis to the required level.  In contrast to the previous
 analysis, the large increase in observation time has provided a
 spectral-quality data set that is well above the detection threshold,
 over a much larger energy range, and that includes improvements in
 procedures used for the analysis of weak, extended sources.  Here, we
 do not assume a particular source morphology and include
 time-dependent optical efficiency corrections to the energy, more
 exclusion regions for background subtraction (for sources that were
 subsequently discovered nearby), and software and lookup-table
 improvements that have reduced systematic errors due to uneven
 exposure.

 \begin{table}
   \begin{center}
     \begin{tabular}{l r@{$\:\pm\:$}l r@{$\:\pm\:$}l r@{$\:\pm\:$}l} 
       \hline\hline
       Region 
       &\multicolumn{2}{c}{F(1-10TeV)} 
       &\multicolumn{2}{c}{\% total} 
       &\multicolumn{2}{c}{$\Gamma$}
       \\
       
       {} 
       & \multicolumn{2}{c}{$\times10^{-12}\UNITS{cm^{-2}\,s^{-1}}$} 
       & \multicolumn{2}{c}{}  
       & \multicolumn{2}{c}{}  \\
       \hline
       Full & 1.63 & 0.16 & \multicolumn{2}{c}{100} & 2.71 & 0.11  \\
       A & 0.25 & 0.04 & 15 & 3 & 2.67 & 0.14\\
       B & 0.11 & 0.02 & 6 & 2 & 2.93 & 0.21 \\
       C & 0.14 & 0.03 & 8 & 2 & 2.86 & 0.27 \\
       \hline\hline
     \end{tabular}
   \end{center}
   \caption{ Integral fluxes of the three test regions \emph{A},
     \emph{B}, and \emph{C} (shown in Figure \ref{fig:VHE}), compared
     with the total integral flux from the full source. The photon
     index $\Gamma$ is derived from a fit to the spectrum for each
     region. Note that region \emph{A} is chosen to correspond with the
     molecular cloud position and radius described in section
     \ref{sec:molecular_clouds}. }  
   \label{tab:intflux}
 \end{table}

Since the emission appears to have a complicated morphology with more
than one peak in the excess image, the possibility that \HESSJ\ is
more than one source was explored. First, three emission peaks were
determined, located at the positions \emph{A}, \emph{B}, and \emph{C}
shown in Figure \ref{fig:VHE}.  Between each pair of peaks, a
one-dimensional slice in the uncorrelated excess image (with a width
of $0.1^\circ$) was made to determine the significance of the ``dip''
between them. In each case, the emission is no more than two standard
deviations from a constant value across the slice.  Furthermore, if
the emission peaks are from multiple sources, one might expect to see
spectral variability across the object, though energy dependent
morphology may also arise from transport and/or energy-loss processes
within a single source. To test this possibility, a spectral analysis
was made at each of the test points with an integration radius of
$0.14^\circ$ for \emph{A} and $0.1^\circ$ for \emph{B} and \emph{C}
(see Table \ref{tab:intflux}).  The spectral indices at each position are
consistent with each other within statistical errors, and also with
the spectrum determined for the entire source region; therefore within
the statistics of the observations, there is no strong evidence to
support the multiple-origin hypothesis.

\section{\XMM\ Observations of the region}

\begin{table*}[h]
  \centering
    \caption{Sources detected using the detection algorithm
    \emph{emldetect}. The parameters given here are for the energy
    range between 0.5 and 10~keV. Col. 2: Name recommended by the
    \XMM\ SOC and the IAU for source detections. Col.\ 3 and 4:
    J2000.0 coordinates. Units of right ascension are hours, minutes,
    and seconds, and units of declination are degrees, arc-minutes,
    and arc-seconds. Col.\ 5: Error on the source position in
    arc-seconds. Col.\ 6: Number of counts in EMOS1 and EMOS2 within a
    10\arcsec\ integration region using events above 0.5~keV. Col.\ 7:
    Statistical significance of the detection derived with
    {\emph{emldetect}}. 8: Source flux above 0.5~keV in
    $10^{-14} \UNITS{erg\,cm^{-2}\,s^{-1}}$. \vspace{0.4cm}}
    \label{tab:xray_ptsrc}
    \begin{tabular}{c | c | c c c | c c c}
      \hline
      Id & XMMU\,J &  RA$_{2000}$ & Dec$_{2000}$ & 2-D Error & Counts
      & Significance & Flux$\cdot10^{-14}$\\
      (1) & (2) & (3) & (4) & (5) & (6) & (7) & (8)\\ \hline
      1 & 174434.5--301522 & \HMS{17}{44}{34.47} & \DMS{-30}{15}{21.9} & 0.6 & 280 & 14.3 & $10.4 \pm 0.7$\\
      2 & 174441.3--301648 & \HMS{17}{44}{41.27} & \DMS{-30}{16}{47.8} & 0.6 & 209 & 12.0 & $8.4 \pm 0.7$ \\
      3 & 174437.6--301812 & \HMS{17}{44}{37.55} & \DMS{-30}{18}{12.3} & 1.3 & 51  & 5.2 & $2.1 \pm 0.4$\\
      4 & 174425.3--302236 & \HMS{17}{44}{25.30} & \DMS{-30}{22}{36.1} & 1.3 & 70 & 7.0 & $3.2 \pm 0.4$\\
      5 & 174458.6--301743 & \HMS{17}{44}{58.59} & \DMS{-30}{17}{43.2} & 0.9 & 95 & 7.5 & $5.2 \pm 0.7$\\
      6 & 174351.0--301709 & \HMS{17}{43}{51.03} & \DMS{-30}{17}{09.0} & 1.0 & 76 & 6.9 & $4.4 \pm 0.6$\\ \hline
    \end{tabular}
\end{table*}


 \begin{figure}
   \centering
   \resizebox{\hsize}{!}{\includegraphics{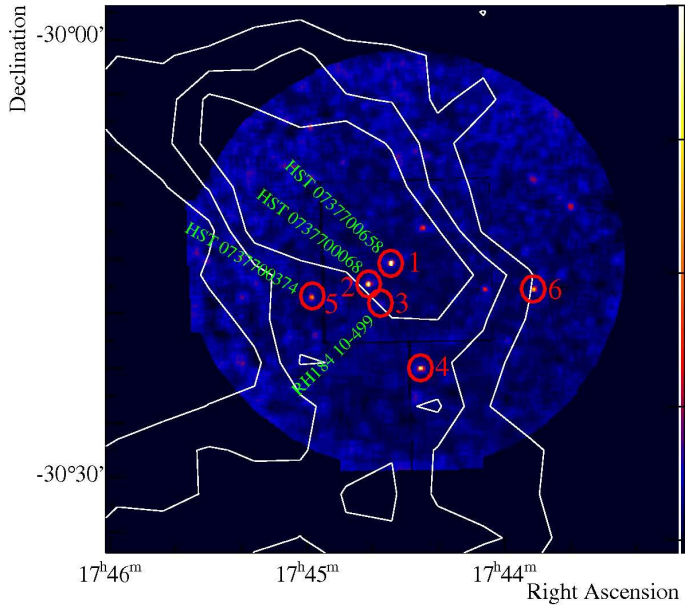}}
   \caption{\label{fig:xmmfov} Point sources detected within the
     \XMM\ exposure (see Table \ref{tab:xray_ptsrc} for
     detailed information). The \HESS\ $4,5,6\sigma$ significance
     contours from Figure \ref{fig:VHE} are
     overlaid in white. Apparently associated stars are labeled. }
 \end{figure}

 \begin{figure}
   \centering
   \resizebox{\hsize}{!}{\includegraphics{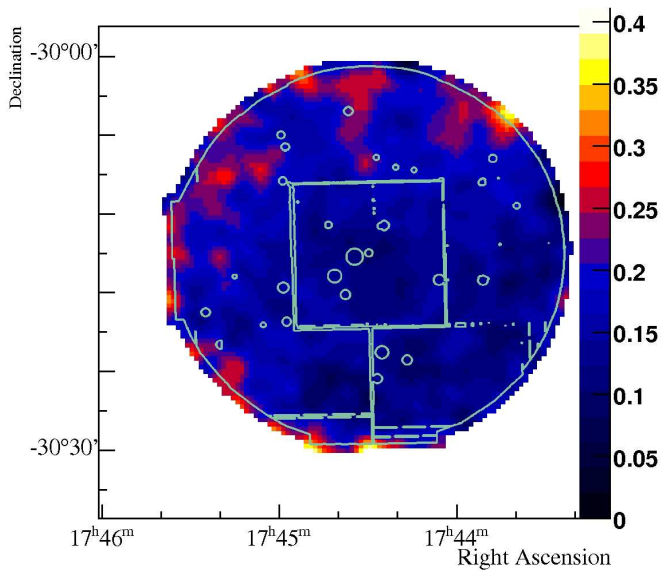}}
   \caption{\label{fig:xraydiffuse} Exposure corrected X-ray image
     smoothed with a $0.01^\circ$ Gaussian filter. All point sources
     above 3$\sigma$ and the borders of each detector chip (shown by
     the contour) have been excluded before smoothing. No evidence for
     diffuse emission is seen. 
}
 \end{figure}

 HESS\,J1745--303 was observed with the \XMM\ X-ray satellite on
 September 18, 2006 for 30~ks in satellite revolution 1241 (ObsID
 0406580201). All X-ray instruments (EPIC MOS~1, MOS~2, and PN) were
 operated in full-frame mode and a medium filter was applied to screen
 out bright optical and UV sources. The calibration, data reduction
 and analysis made use of the \XMM\ Science Analysis Software (SAS),
 version 7.0, together with the Extended Source Analysis Software
 package (\emph{XMM}-ESAS), version
 1.0~\cite{snowden04:ESAS}. Following standard data reduction and
 calibration procedures, the data set was cleaned from temporally
 occurring background caused by soft proton flares. The resulting
 observation time amounts to 14.4~ks of useful
 data. Figure~\ref{fig:xmmfov} shows an adaptively smoothed count map
 of the region surrounding \HESSJ\ using events above 0.5~keV detected
 with either the MOS~1 or MOS~2 detector of \XMM. The white contours
 indicate the \HESS\ VHE gamma-ray significance contours at $4\sigma$,
 $5\sigma$ and $6\sigma$. Six X-ray sources are apparent (labeled in
 red), as determined by the standard \XMM\ source detection algorithm
 {\emph{emldetect}} in the energy band from 0.5--10~keV, as well as in
 sub-intervals from 0.5--2~keV, 2.0--4.5~keV and from
 4.5--10~keV. Table~\ref{tab:xray_ptsrc} summarizes the sources
 detected above 0.5~keV. For all these sources the algorithm also
 attempts to determine a source extension by fitting a Gaussian model
 to the data. All six sources were found to be consistent with a
 point-source. Several of these X-ray sources coincide with stars
 known in the optical (shown in green in Figure~\ref{fig:xmmfov}).
 Sources 1, 2, and 5 were found to be positionally coincident with HST
 optical guide stars as shown in Figure~\ref{fig:xmmfov}, source 3 was
 found to coincide with an M3-star (RHI84\, 10--499) and for sources 4
 and 6 a catalog search did not yield any obvious counterparts in
 other wavebands. As all of the detected X-ray sources are point-like
 and rather faint with fluxes around or below
 $10^{-14}\UNITS{erg\,cm^{-2}\,s^{-1}}$, assuming a similar energy
 flux in X-rays and gamma rays, it seems unlikely that any of these
 sources are connected to the bright extended VHE gamma-ray source
 HESS\,J1745--303, which is extended and has an energy flux of
 $(2.00\pm0.18)\times 10^{-11}\UNITS{erg\,cm^{-2}\,s^{-1}}$ above
 200~GeV.  In addition to the search for point-sources an analysis
 sensitive to diffuse X-ray emission has been performed. To that end,
 sources detected above a significance level of 3$\sigma$ were
 excluded from the raw counts map and the exposure map. The maps were
 then smoothed with a Gaussian of width 0.01$^{\circ}$ and the ratio
 taken to produce the resulting smoothed, exposure-corrected counts
 map as shown in Figure~\ref{fig:xraydiffuse}.  Diffuse emission at a
 level similar to or above the level of the detected point-sources
 should show up in this method.  However, no sign of such a diffuse
 emission is detected in the whole field of view and we derive a 99\%
 confidence limit on the flux level of the diffuse emission in region
 \emph{A} of $4.5 \times 10^{-13}\UNITS{erg\,cm^{-2}\,s^{-1}}$, using
 the rest of the field-of-view to determine the background, and $7.1
 \times 10^{-13} \UNITS{erg\,cm^{-2}\,s^{-1}}$ using only a strip to
 the west of region \emph{A} as background.


\section{Possible associations} \label{sec:assoc}

To look for possible associations, standard catalogs of sources
thought to be associated with VHE emission were searched, including
high-spin-down flux\footnote{High spin-down flux ($\dot E/D^2$)
pulsars are thought to be likely VHE emission candidates if the
conversion efficiency from spin-down power to VHE emission is around
1\% \cite[e.g.][]{carrigan07}.} pulsars \cite{ATNF}, SNRs
\cite{green04:SNRs}, Wolf-Rayet stars
\cite{hucht01:_catal_galac_wolf_rayet}, high-mass X-ray binaries
\cite{liu06:HMXB}, INTEGRAL sources \cite{INTEGRAL:3IBIS}, and HII
regions.  Additionally, public survey data from ROSAT \cite{ROSAT},
ASCA \cite{ASCA}, and the VLA
\cite{larosa00:_wide_field_centim_vla_image} were searched for
possible un-cataloged counterparts. The most likely candidates found
in this search are discussed here.

\subsection{\SNRa\  and Molecular Clouds} \label{sec:molecular_clouds}

\begin{figure}
  \centering
  \resizebox{\hsize}{!}{\includegraphics{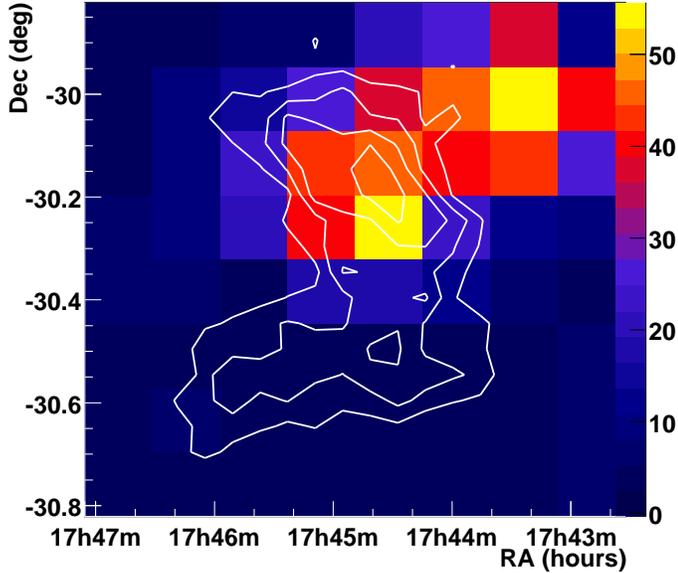}}
  \caption{\label{fig:CO} Velocity-integrated ($-100$ to
    $-60\,\KMPERSEC$) map of \CO\
    \cite{bitran97:_large_scale_co_survey} emission showing the
    molecular cloud coincident with the shell of \SNRa, with overlaid
    \HESS\ $4\sigma$ to $7\sigma$ significance contours.}
\end{figure}

\SNRa\ was identified as an SNR by \citet{downes79:radio_gc} using the
Westerbork Synthesis Radio Telescope and the Bonn 100-m telescope. VLA
observations \cite{uchida92:_h_i_absor_line_study} confirmed the
presence of the non-thermal shell and HI absorption showed that its
probable location is within a few hundred parsecs of the Galactic
center. At an estimated distance of 7.6~kpc, the $12'$ radius of the
remnant corresponds to 26.5~pc, with an estimated age of $\ge 10^4$ yr
(a middle aged SNR). Observations of the \CO\ emission line
(J=1$\to$0) with the Bell Laboratories telescope
\cite{uchida92:_dense_molec_ring_surroun_nonth} revealed a dense ring
of matter surrounding the shell. The radial velocity dispersion of
this super-shell between $-60$ and $-190\,\KMPERSEC$ agrees well with
a location of the remnant close to the Galactic center. This
super-shell could have been produced by the combined stellar winds of
$\sim$200 O-type stars concentrated in this region , which is very
probably the birth place of the remnant
\cite{uchida92:_dense_molec_ring_surroun_nonth}.  X-ray observations
of the remnant with ASCA showed no evidence for a shell in the energy
range 3.2--10.0\,keV, detecting only thermal diffuse emission from the
central region \cite{bamba00:_g359}.

\citet{uchida92:_dense_molec_ring_surroun_nonth} observed a
correlation in intensity between the non-thermal radio emission of the
remnant and the CO emission and suggested an association of the
two. There are no sharp gradients observed in the velocity
distribution, the absence of which could be explained by the presence
of magnetic precursors or by a previous acceleration of the shell by
stellar winds or old supernovae.  Another indication of the
interaction of the remnant with the surrounding medium is the presence
of maser emission spots near the edge of the shell, revealed by VLA
observations \cite{yusef-zadeh95:_shock_excit_oh_maser_emiss}. OH
masers at 1720 MHz are believed to be produced by collisional pumping
behind shocks and are therefore a good indicator of shocked clouds
\cite{elitzur76:_inver_of_oh_mhz_line}. Although the mean velocity of
the masers at around $-5\,\KMPERSEC$ is significantly shifted from the
velocity range of the \CO\ shell ([$-60$,$-190$] $\KMPERSEC$), a
random coincidence of the shell with the maser positions seems
unlikely. \citet{yusef-zadeh95:_shock_excit_oh_maser_emiss} discussed
this velocity discrepancy: the two velocities could be reconciled if
the shocked cloud was carried away by the shock itself, which could
have redirected the radial velocity component. The specific conditions
required to allow a population inversion (and thus maser emission) may
also explain this discrepancy.  Moreover, the association between the
OH masers, the SNR and the \CO\ cloud is supported by the fact that
the maser distribution shows a good correlation with the \CO\ emission
maximum and the non-thermal radio emission from the western part of
the shell.

We have used \CO\ (J=1$\to$0) data from the Cerro Tololo
Inter-American Observatory (Chile) to map the matter distribution in
the western part of the remnant
\cite{bitran97:_large_scale_co_survey}. Figure \ref{fig:CO} is the
\CO\ map integrated in velocity between $-100 \,\KMPERSEC$ and
$-60\,\KMPERSEC$. This velocity range corresponds to the western part
of the super-shell surrounding the remnant and is potentially
associated with the remnant. Two overlapping clouds are present in
this region. The cloud shown in Figure \ref{fig:CO} contains a
component that is partially coincident with \HESSJ.

Due to the relatively high magnetic field, (0.2--0.6\,mG, as measured
from the Zeeman splitting of maser lines
\citep{robinson96:_mhz_oh_maser_emiss_snr_g359}, the cooling time of
ultra-relativistic electrons would be much shorter than the age of the
remnant, and therefore an electron production scenario seems unlikely
as an explanation of the VHE emission. A possible explanation is that
a part of the VHE gamma-ray source comes instead from hadronic
cosmic-ray interactions in this cloud, producing neutral pions that
decay into two gamma rays.  Assuming a value of the ratio $X =
N_{\mathrm{H_2}}/W_{\mathrm{CO}}$ of
$1.8\times10^{20}\UNITS{cm^{-2}\,K^{-1}\,km^{-1}\,s^{-1}}$
\cite{dame01:_milky_way_in_molec_cloud}, we estimate the H$_2$ mass in
this cloud to be $5\times10^4 M_\odot$ with a density of $5\times
10^3\UNITS{cm^{-3}}$, assuming spherical symmetry.  The value of the
ratio $X$ used is a mean galactic value which may not be relevant for
the inner part of the galaxy, therefore the cloud mass may be
overestimated. A spectral analysis of the part of \HESSJ\ coincident
with the cloud (region \emph{A}) indicates that approximately 15\% of
the observed flux comes from this region (see Table
\ref{tab:intflux}).

Since hadronic interactions lead to the production of gamma rays with
energies typically a factor of 10 below the primary energy, the energy
of interacting protons ($W$) required to generate the observed flux of
VHE $\gamma$-rays between 300 GeV and 40~TeV can be estimated in the
corresponding energy range of approximately 3--400\,TeV to be: $W =
t_{\mathrm{pp}\rightarrow\pi^{0}} \cdot L_{\gamma}$, where
$t_{\mathrm{pp}\rightarrow\pi^{0}}$ is the characteristic cooling time
of protons through the \UPDATED{$\pi^{0}$} production channel and
$L_{\gamma}$ is the gamma-ray luminosity between 300~GeV and 40~TeV
\cite{aharonian96:_emiss_of_pi_decay_gamma}. Assuming that the proton
energy distribution follows a power law with the same index as the
gamma rays over the full relativistic range, we can extrapolate this
distribution down to 1\,GeV. Assuming also that the accelerated proton
density is uniform in the whole remnant, it corresponds to a fraction
of \UPDATED{$\sim$32\%} of the mechanical explosion energy of
10$^{51}$~erg of the remnant. This estimate suffers from large
uncertainties, mainly from the cloud mass estimation, the fraction of
HESS J1745-303 involved in this association, and the explosion
energy. However, it is interesting that we obtain an estimate that is
comparable to the theoretical espectation of $\sim$10\%.

\subsection{\EGRSRC}

Most of the emission associated with \HESSJ\ lies within the 95\%
error-circle of the unidentified EGRET source \EGRSRC\ (marked with a
dotted circle in Figure \ref{fig:VHE}), which has an integral flux in
the 100 MeV--10 GeV energy band of $(63.9
\pm7.1)\times10^{-8}\UNITS{cm^{-2}\,s^{-1}}$ with a photon index of
$\Gamma_\mathrm{EGRET} = 2.17 \pm 0.08$ \cite{EGRET:3EG}.
Extrapolating this flux to the VHE range, we find an expected integral
flux in the range 1--10 TeV of
$(3^{+4}_{-1})\times10^{-11}\UNITS{cm^{-2}\,s^{-1}}$, which is higher
than the integral flux observed by \HESS\ over the same range (see
Figure \ref{fig:spect}).  Fitting a power law with an exponential
cutoff to the \HESS\ spectrum combined with the EGRET flux point, we
find a cutoff energy of $\sim$$(3.0\pm1.5)\UNITS{TeV}$.

A study of the long-term variability of EGRET sources made by
\citet{torres01:_variab_analy_of_low_latit} shows that \EGRSRC\ is
variable (on year timescales), with an average statistical index of
variability over $3\sigma$ higher than that expected from pulsars
(which are considered a non-variable source used as a reference for
systematic variability of the instrument). If the EGRET source is
truly variable, it is unlikely to be associated with the extended
emission seen in \HESSJ, where no variability would be expected.


\subsection{ \PSRa,  \PSRb, SLX~1744-299 }

Two cataloged pulsars lie within or near \HESSJ: \PSRa\ and \PSRb\
(see Figure \ref{fig:VHE}). Energetic pulsars driving pulsar wind
nebulae are known to produce VHE emission that may be asymmetric or
offset from the pulsar position
\cite[e.g.][]{HESS:VelaX,HESS:Kookaburra,HESS:J1825,HESS:MSH1552}. In
several cases, the PWN candidate has been first identified in the VHE
energy range and subsequently confirmed with X-ray measurements
\cite[e.g.][]{HESS:2PSR,hinton07:hessj1718_XMM}.

\PSRa\ (also known as PSR~J1745$-$3040) is a rather old pulsar (546
kyr) located near the southern edge of the region of significant
emission in \HESSJ, and has a spin-down flux
$\dot{E}/D^2=2\times10^{33}\UNITS{erg\,s^{-1}\,kpc^{-2}}$ \cite{ATNF},
requiring a conversion efficiency from rotational kinetic energy to
gamma-ray emission of approximately 32\% to produce the entire VHE
emission. Such high apparent efficiencies are possible if the
spin-down flux was much higher in the past and the particle cooling
times are comparable to or much shorter than the pulsar age. If \PSRa\
powers only the fraction of \HESSJ\ enclosed by region \emph{B}, then
the required conversion efficiency would be only 2\%, which is not
unreasonable compared with other known VHE PWNe.

\PSRb\ is located approximately a half-degree east of \HESSJ\ and is
associated with the bright X-ray and radio feature G359.23$-$0.82,
also known as ``the Mouse'' \cite{gaensler04:_mouse_soared} (seen in
the radio image in Figure \ref{fig:VHE}), which is believed to be a
bow-shock PWN with a trailing tail caused by the reverse termination
shock. Given the proper velocity, estimated to be
$\sim600\UNITS{km\,s^{-1}}$ \cite{gaensler04:_mouse_soared}, distance
$D \simeq 2.08\UNITS{kpc}$ and age (25.5 kyr) \cite{ATNF}, one can
extrapolate that the pulsar would have moved approximately
$0.43^\circ$ from its original position, placing it close to \HESSJ\
if the direction of motion is along the tail of the ``Mouse'' (see
dotted line in Figure \ref{fig:VHE}, Bottom).  The relatively high
spin-down flux of \PSRb\
($\dot{E}/D^2=4\times10^{35}\mathrm{erg\,s^{-1}\,kpc^{-2}}$) would
imply a 0.2\% conversion efficiency to explain the entire VHE
emission, or 0.02\% for only region \emph{A}. In this case, the PWN
would have to be extremely offset and asymmetric.

Located near \PSRb\ are the ultra-compact X-ray binary SLX~1744-299
\cite{zand07:1744_299}, and the X-ray burster $1744-300$
\cite{skinner90}, which are not generally expected to produce offset
or extended emission and are thus not considered probable counterparts
to \HESSJ.

\section{Discussion}

Due to the positional coincidence and plausible energetics, at least
part of the emission of \HESSJ\ (in the region labeled \emph{A} in
Figure \ref{fig:VHE}) may well be associated with the interaction of
the shell of \SNRa\ with a molecular cloud. This scenario fits
particularly well within the context of theoretical predictions for
VHE gamma-ray emission from SNRs embedded in dense media.
\citet{gabici07} show that for a SNR of approximately the same age as
\SNRa\ in this context, one would expect significant TeV emission that
peaks around 1 TeV.  Assuming a similar supernova energy output of
$10^{51}\UNITS{erg}$ and scaling their theoretical model for a cloud
at 30\,pc away from an SNR shell by $M_\mathrm{cl}/D^2$, where
$M_\mathrm{cl}$ is the mass of the cloud near \SNRa\ and $D$ is its
distance, we find that the predicted flux at 1 TeV of
$2.8\times10^{-12}\UNITS{TeV^{-1}\,cm^{-2}\,s^{-1}}$ matches well with
the \HESSJ\ emission around region \emph{A},
$F_A(1\UNITS{TeV})=(4.4\pm0.5)\times10^{-12}\UNITS{TeV^{-1}\,cm^{-2}\,s^{-1}}$,
as does the soft $\Gamma\simeq3.0$ spectral index.

However, since the VHE emission extends beyond region \emph{A} (and
the dense target material does not), the SNR/molecular-cloud scenario
is not sufficient for describing the entire VHE source. Since there is
as yet no statistically significant separation between the various
parts of the emission region, an explanation for the entire source
remains complicated and the possibility of source confusion still
remains.  In particular, \PSRa\ is energetic enough to power a PWN in
the part of \HESSJ\ surrounding it (region \emph{B}), and though an
unlikely candidate due to its significant offset, \PSRb\ is in
principle powerful enough to power the entire VHE source. The lack of
significant spectral variability across the emission region further
complicates the identification of counterparts.  It is important to
note that many VHE sources and source classes have spectral indices in
the range 2.1--2.7, therefore given the statistics, it would be likely
impossible to disentangle a superposition of three randomly chosen VHE
sources from their spectra alone.

Furthermore, the lack of a significant extended source in the \XMM\
data is not without precedent---several other TeV sources, such as
HESS$-$J1303-631 \cite{HESS:J1303}, other unidentified TeV sources
\cite{HESS:unidentified}, have so far no identified X-ray counterparts
The lack of non-thermal X-ray or radio emission combined with the
relatively high magnetic field around \SNRa\ further supports the
hadronic scenario for VHE gamma-ray production for the region
associated with the interaction of the SNR shock with the target
material in the molecular cloud. However, even in a hadronic scenario,
some longer-wavelength emission would be expected due to secondary
electrons produced in the interactions.

An association of \EGRSRC\ with part or all of \HESSJ\ is also
plausible from an energetic standpoint
\citep[e.g.][]{funk07:_gev_tev_connec_galac}, however since the size
of the EGRET error circle is larger than the VHE emission, the
position may not correspond with the VHE source and further
localization is not currently possible.
\citet{aharonian96:_emiss_of_pi_decay_gamma} show that the observed
gamma-ray flux from a hadronic source is proportional to
$E_\gamma^{-(\Gamma_p+\delta)}$, where $\Gamma_p$ is the proton index at
the source and $\delta$ is the index of the diffusion coefficient
(typically $0.3-0.6$), allowing spectra that are quite soft in the TeV
energy range, and which may be different from the slopes in other
energy bands \cite{torres03:_super_remnan_and_gamma_ray_sourc}.
Therefore for an $E_p^{-2.0}$ source spectrum, it is possible to
reproduce both the hard ($\Gamma=2.17$) EGRET spectrum and the softer
$\Gamma=\HESSINDEX$ TeV spectrum. If we assume \EGRSRC\ is associated
only with the SNR/molecular-cloud interaction, then the observed flux
is significantly higher than that predicted by the \citet{gabici07}
model, however given the uncertainties in position, the EGRET flux may
also contain contributions from the SNR shell itself (where gamma rays
may be produced by e.g. inverse-Compton scattering of lower energy
photons by high-energy electrons accelerated in the shock), or from
other sources in the region, which could explain this.

Further multi-wavelength observations in X-rays and GeV gamma rays
(e.g. from observatories such as XMM, Suzaku and GLAST) of the entire
region spanned by \HESSJ\ as well as deeper VHE exposures will be
needed to disentangle the emission possibilities and to find more
definitive counterparts.

\begin{acknowledgements}

  The support of the Namibian authorities and of the University of
  Namibia in facilitating the construction and operation of \HESS\ is
  gratefully acknowledged, as is the support by the German Ministry
  for Education and Research (BMBF), the Max Planck Society, the
  French Ministry for Research, the CNRS-IN2P3 and the Astroparticle
  Interdisciplinary Programme of the CNRS, the U.K. Science and
  Technology Facilities Council (STFC), the IPNP of the Charles
  University, the Polish Ministry of Science and Higher Education, the
  South African Department of Science and Technology and National
  Research Foundation, and by the University of Namibia. We appreciate
  the excellent work of the technical support staff in Berlin, Durham,
  Hamburg, Heidelberg, Palaiseau, Paris, Saclay, and in Namibia in the
  construction and operation of the equipment.

  We would like to further thank Thomas Dame for providing us with the
  CO data. This research has made use of the SIMBAD database, operated
  at CDS, Strasbourg, France and the ROSAT Data Archive of the
  Max-Planck-Institut f\"ur extraterrestrische Physik (MPE) at
  Garching, Germany.

\end{acknowledgements}

\bibliographystyle{aa} \bibliography{kpk}

\begin{thebibliography}{55}
\expandafter\ifx\csname natexlab\endcsname\relax\def\natexlab#1{#1}\fi

\bibitem[{{Aharonian} {et~al.}(2005{\natexlab{a}}){Aharonian}, {Akhperjanian},
  {Aye}, {Bazer-Bachi}, {Beilicke}, {Benbow}, {Berge}, {Berghaus},
  {Bernl{\"o}hr}, {Boisson}, {Bolz}, {Borgmeier}, {Braun}, {Breitling},
  {Brown}, {Bussons Gordo}, {Chadwick}, {Chounet}, {Cornils}, {Costamante},
  {Degrange}, {Djannati-Ata{\"i}}, {O'C.~Drury}, {Dubus}, {Ergin}, {Espigat},
  {Feinstein}, {Fleury}, {Fontaine}, {Funk}, {Gallant}, {Giebels}, {Gillessen},
  {Goret}, {Hadjichristidis}, {Hauser}, {Heinzelmann}, {Henri}, {Hermann},
  {Hinton}, {Hofmann}, {Holleran}, {Horns}, {de Jager}, {Jung}, {Kh{\'e}lifi},
  {Komin}, {Konopelko}, {Latham}, {Le Gallou}, {Lemi{\`e}re}, {Lemoine},
  {Leroy}, {Lohse}, {Marcowith}, {Masterson}, {McComb}, {de Naurois}, {Nolan},
  {Noutsos}, {Orford}, {Osborne}, {Ouchrif}, {Panter}, {Pelletier}, {Pita},
  {P{\"u}hlhofer}, {Punch}, {Raubenheimer}, {Raue}, {Raux}, {Rayner},
  {Redondo}, {Reimer}, {Reimer}, {Ripken}, {Rob}, {Rolland}, {Rowell},
  {Sahakian}, {Saug{\'e}}, {Schlenker}, {Schlickeiser}, {Schuster}, {Schwanke},
  {Siewert}, {Sol}, {Steenkamp}, {Stegmann}, {Tavernet}, {Terrier},
  {Th{\'e}oret}, {Tluczykont}, {Vasileiadis}, {Venter}, {Vincent}, {Visser},
  {V{\"o}lk}, \& {Wagner}}]{HESS:G09}
{Aharonian}, F., {Akhperjanian}, A.~G., {Aye}, K.-M., {et~al.}
  2005{\natexlab{a}}, \aap, 432, L25

\bibitem[{{Aharonian} {et~al.}(2005{\natexlab{b}}){Aharonian}, {Akhperjanian},
  {Aye}, {Bazer-Bachi}, {Beilicke}, {Benbow}, {Berge}, {Berghaus},
  {Bernl{\"o}hr}, {Boisson}, {Bolz}, {Braun}, {Breitling}, {Brown}, {Bussons
  Gordo}, {Chadwick}, {Chounet}, {Cornils}, {Costamante}, {Degrange},
  {Djannati-Ata{\"i}}, {O'C.~Drury}, {Dubus}, {Emmanoulopoulos}, {Espigat},
  {Feinstein}, {Fleury}, {Fontaine}, {Fuchs}, {Funk}, {Gallant}, {Giebels},
  {Gillessen}, {Glicenstein}, {Goret}, {Hadjichristidis}, {Hauser},
  {Heinzelmann}, {Henri}, {Hermann}, {Hinton}, {Hofmann}, {Holleran}, {Horns},
  {de Jager}, {Kh{\'e}lifi}, {Komin}, {Konopelko}, {Latham}, {Le Gallou},
  {Lemi{\`e}re}, {Lemoine-Goumard}, {Leroy}, {Lohse}, {Martineau-Huynh},
  {Marcowith}, {Masterson}, {McComb}, {de Naurois}, {Nolan}, {Noutsos},
  {Orford}, {Osborne}, {Ouchrif}, {Panter}, {Pelletier}, {Pita},
  {P{\"u}hlhofer}, {Punch}, {Raubenheimer}, {Raue}, {Raux}, {Rayner},
  {Redondo}, {Reimer}, {Reimer}, {Ripken}, {Rob}, {Rolland}, {Rowell},
  {Sahakian}, {Saug{\'e}}, {Schlenker}, {Schlickeiser}, {Schuster}, {Schwanke},
  {Siewert}, {Sol}, {Steenkamp}, {Stegmann}, {Tavernet}, {Terrier},
  {Th{\'e}oret}, {Tluczykont}, {Vasileiadis}, {Venter}, {Vincent}, {V{\"o}lk},
  \& {Wagner}}]{HESS:MSH1552}
{Aharonian}, F., {Akhperjanian}, A.~G., {Aye}, K.-M., {et~al.}
  2005{\natexlab{b}}, \aap, 435, L17

\bibitem[{{Aharonian} {et~al.}(2005{\natexlab{c}}){Aharonian}, {Akhperjanian},
  {Aye}, {Bazer-Bachi}, {Beilicke}, {Benbow}, {Berge}, {Berghaus},
  {Bernl{\"o}hr}, {Boisson}, {Bolz}, {Braun}, {Breitling}, {Brown}, {Bussons
  Gordo}, {Chadwick}, {Chounet}, {Cornils}, {Costamante}, {Degrange},
  {Djannati-Ata{\"i}}, {O'C.~Drury}, {Dubus}, {Emmanoulopoulos}, {Espigat},
  {Feinstein}, {Fleury}, {Fontaine}, {Fuchs}, {Funk}, {Gallant}, {Giebels},
  {Gillessen}, {Glicenstein}, {Goret}, {Hadjichristidis}, {Hauser},
  {Heinzelmann}, {Henri}, {Hermann}, {Hinton}, {Hofmann}, {Holleran}, {Horns},
  {de Jager}, {Kh{\'e}lifi}, {Komin}, {Konopelko}, {Latham}, {Le Gallou},
  {Lemi{\`e}re}, {Lemoine-Goumard}, {Leroy}, {Lohse}, {Martineau-Huynh},
  {Marcowith}, {Masterson}, {McComb}, {de Naurois}, {Nolan}, {Noutsos},
  {Orford}, {Osborne}, {Ouchrif}, {Panter}, {Pelletier}, {Pita},
  {P{\"u}hlhofer}, {Punch}, {Raubenheimer}, {Raue}, {Raux}, {Rayner},
  {Redondo}, {Reimer}, {Reimer}, {Ripken}, {Rob}, {Rolland}, {Rowell},
  {Sahakian}, {Saug{\'e}}, {Schlenker}, {Schlickeiser}, {Schuster}, {Schwanke},
  {Siewert}, {Sol}, {Steenkamp}, {Stegmann}, {Tavernet}, {Terrier},
  {Th{\'e}oret}, {Tluczykont}, {van der Walt}, {Vasileiadis}, {Venter},
  {Vincent}, {V{\"o}lk}, \& {Wagner}}]{HESS:J1303}
{Aharonian}, F., {Akhperjanian}, A.~G., {Aye}, K.-M., {et~al.}
  2005{\natexlab{c}}, \aap, 439, 1013

\bibitem[{{Aharonian} {et~al.}(2004){Aharonian}, {Akhperjanian}, {Aye},
  {Bazer-Bachi}, {Beilicke}, {Benbow}, {Berge}, {Berghaus}, {Bernl{\"o}hr},
  {Bolz}, {Boisson}, {Borgmeier}, {Breitling}, {Brown}, {Bussons Gordo},
  {Chadwick}, {Chitnis}, {Chounet}, {Cornils}, {Costamante}, {Degrange},
  {Djannati-Ata{\"i}}, {O'C.~Drury}, {Ergin}, {Espigat}, {Feinstein}, {Fleury},
  {Fontaine}, {Funk}, {Gallant}, {Giebels}, {Gillessen}, {Goret}, {Guy},
  {Hadjichristidis}, {Hauser}, {Heinzelmann}, {Henri}, {Hermann}, {Hinton},
  {Hofmann}, {Holleran}, {Horns}, {de Jager}, {Jung}, {Kh{\'e}lifi}, {Komin},
  {Konopelko}, {Latham}, {Le Gallou}, {Lemoine}, {Lemi{\`e}re}, {Leroy},
  {Lohse}, {Marcowith}, {Masterson}, {McComb}, {de Naurois}, {Nolan},
  {Noutsos}, {Orford}, {Osborne}, {Ouchrif}, {Panter}, {Pelletier}, {Pita},
  {Pohl}, {P{\"u}hlhofer}, {Punch}, {Raubenheimer}, {Raue}, {Raux}, {Rayner},
  {Redondo}, {Reimer}, {Reimer}, {Ripken}, {Rivoal}, {Rob}, {Rolland},
  {Rowell}, {Sahakian}, {Saug{\'e}}, {Schlenker}, {Schlickeiser}, {Schuster},
  {Schwanke}, {Siewert}, {Sol}, {Steenkamp}, {Stegmann}, {Tavernet},
  {Th{\'e}oret}, {Tluczykont}, {van der Walt}, {Vasileiadis}, {Vincent},
  {Visser}, {V{\"o}lk}, \& {Wagner}}]{HESS:GC}
{Aharonian}, F., {Akhperjanian}, A.~G., {Aye}, K.-M., {et~al.} 2004, \aap, 425,
  L13

\bibitem[{{Aharonian} {et~al.}(2007){Aharonian}, {Akhperjanian}, {Bazer-Bachi},
  {Behera}, {Beilicke}, {Benbow}, {Berge}, {Bernl{\"o}hr}, {Boisson}, {Bolz},
  {Borrel}, {Braun}, {Brion}, {Brown}, {B{\"u}hler}, {B{\"u}sching},
  {Boutelier}, {Carrigan}, {Chadwick}, {Chounet}, {Coignet}, {Cornils},
  {Costamante}, {Degrange}, {Dickinson}, {Djannati-Ata{\"i}}, {Domainko},
  {Drury}, {Dubus}, {Egberts}, {Emmanoulopoulos}, {Espigat}, {Farnier},
  {Feinstein}, {Fiasson}, {F{\"o}rster}, {Fontaine}, {Funk}, {Funk},
  {F{\"u}{\ss}ling}, {Gallant}, {Giebels}, {Glicenstein}, {Gl{\"u}ck}, {Goret},
  {Hadjichristidis}, {Hauser}, {Hauser}, {Heinzelmann}, {Henri}, {Hermann},
  {Hinton}, {Hoffmann}, {Hofmann}, {Holleran}, {Hoppe}, {Horns},
  {Jacholkowska}, {de Jager}, {Kendziorra}, {Kerschhaggl}, {Kh{\'e}lifi},
  {Komin}, {Kosack}, {Lamanna}, {Latham}, {Le Gallou}, {Lemi{\`e}re},
  {Lemoine-Goumard}, {Lohse}, {Martin}, {Martineau-Huynh}, {Marcowith},
  {Masterson}, {Maurin}, {McComb}, {Moulin}, {de Naurois}, {Nedbal}, {Nolan},
  {Noutsos}, {Olive}, {Orford}, {Osborne}, {Panter}, {Pedaletti}, {Pelletier},
  {Petrucci}, {Pita}, {P{\"u}hlhofer}, {Punch}, {Ranchon}, {Raubenheimer},
  {Raue}, {Rayner}, {Ripken}, {Rob}, {Rolland}, {Rosier-Lees}, {Rowell},
  {Ruppel}, {Sahakian}, {Santangelo}, {Saug{\'e}}, {Schlenker}, {Schlickeiser},
  {Schr{\"o}der}, {Schwanke}, {Schwarzburg}, {Schwemmer}, {Shalchi}, {Sol},
  {Spangler}, {Steenkamp}, {Stegmann}, {Superina}, {Tam}, {Tavernet},
  {Terrier}, {Tluczykont}, {van Eldik}, {Vasileiadis}, {Venter}, {Vialle},
  {Vincent}, {V{\"o}lk}, {Wagner}, \& {Ward}}]{HESS:2PSR}
{Aharonian}, F., {Akhperjanian}, A.~G., {Bazer-Bachi}, A.~R., {et~al.} 2007,
  \aap, 472, 489

\bibitem[{{Aharonian} {et~al.}(2008){Aharonian}, {Akhperjanian}, {Bazer-Bachi},
  {Behera}, {Beilicke}, {Benbow}, {Berge}, {Bernl{\"o}hr}, {Boisson}, {Bolz},
  {Borrel}, {Braun}, {Brion}, {Brown}, {B{\"u}hler}, {B{\"u}sching},
  {Boutelier}, {Carrigan}, {Chadwick}, {Chounet}, {Coignet}, {Cornils},
  {Costamante}, {Degrange}, {Dickinson}, {Djannati-Ata{\"i}}, {Domainko},
  {Drury}, {Dubus}, {Egberts}, {Emmanoulopoulos}, {Espigat}, {Farnier},
  {Feinstein}, {Fiasson}, {F{\"o}rster}, {Fontaine}, {Funk}, {Funk},
  {F{\"u}{\ss}ling}, {Gallant}, {Giebels}, {Glicenstein}, {Gl{\"u}ck}, {Goret},
  {Hadjichristidis}, {Hauser}, {Hauser}, {Heinzelmann}, {Henri}, {Hermann},
  {Hinton}, {Hoffmann}, {Hofmann}, {Holleran}, {Hoppe}, {Horns},
  {Jacholkowska}, {de Jager}, {Kendziorra}, {Kerschhaggl}, {Kh{\'e}lifi},
  {Komin}, {Kosack}, {Lamanna}, {Latham}, {Le Gallou}, {Lemi{\`e}re},
  {Lemoine-Goumard}, {Lohse}, {Martin}, {Martineau-Huynh}, {Marcowith},
  {Masterson}, {Maurin}, {McComb}, {Moulin}, {de Naurois}, {Nedbal}, {Nolan},
  {Noutsos}, {Olive}, {Orford}, {Osborne}, {Panter}, {Pedaletti}, {Pelletier},
  {Petrucci}, {Pita}, {P{\"u}hlhofer}, {Punch}, {Ranchon}, {Raubenheimer},
  {Raue}, {Rayner}, {Ripken}, {Rob}, {Rolland}, {Rosier-Lees}, {Rowell},
  {Ruppel}, {Sahakian}, {Santangelo}, {Saug{\'e}}, {Schlenker}, {Schlickeiser},
  {Schr{\"o}der}, {Schwanke}, {Schwarzburg}, {Schwemmer}, {Shalchi}, {Sol},
  {Spangler}, {Steenkamp}, {Stegmann}, {Superina}, {Tam}, {Tavernet},
  {Terrier}, {Tluczykont}, {van Eldik}, {Vasileiadis}, {Venter}, {Vialle},
  {Vincent}, {V{\"o}lk}, {Wagner}, \& {Ward}}]{HESS:unidentified}
{Aharonian}, F., {Akhperjanian}, A.~G., {Bazer-Bachi}, A.~R., {et~al.} 2008,
  \aa, 477

\bibitem[{{Aharonian} {et~al.}(2006{\natexlab{a}}){Aharonian}, {Akhperjanian},
  {Bazer-Bachi}, {Beilicke}, {Benbow}, {Berge}, {Bernl{\"o}hr}, {Boisson},
  {Bolz}, {Borrel}, {Braun}, {Breitling}, {Brown}, {B{\"u}hler},
  {B{\"u}sching}, {Carrigan}, {Chadwick}, {Chounet}, {Cornils}, {Costamante},
  {Degrange}, {Dickinson}, {Djannati-Ata{\"i}}, {O'C.~Drury}, {Dubus},
  {Egberts}, {Emmanoulopoulos}, {Espigat}, {Feinstein}, {Ferrero}, {Fiasson},
  {Fontaine}, {Funk}, {Funk}, {Gallant}, {Giebels}, {Glicenstein}, {Goret},
  {Hadjichristidis}, {Hauser}, {Hauser}, {Heinzelmann}, {Henri}, {Hermann},
  {Hinton}, {Hofmann}, {Holleran}, {Horns}, {Jacholkowska}, {de Jager},
  {Kh{\'e}lifi}, {Komin}, {Konopelko}, {Kosack}, {Latham}, {Le Gallou},
  {Lemi{\`e}re}, {Lemoine-Goumard}, {Lohse}, {Martin}, {Martineau-Huynh},
  {Marcowith}, {Masterson}, {McComb}, {de Naurois}, {Nedbal}, {Nolan},
  {Noutsos}, {Orford}, {Osborne}, {Ouchrif}, {Panter}, {Pelletier}, {Pita},
  {P{\"u}hlhofer}, {Punch}, {Raubenheimer}, {Raue}, {Rayner}, {Reimer},
  {Reimer}, {Ripken}, {Rob}, {Rolland}, {Rowell}, {Sahakian}, {Saug{\'e}},
  {Schlenker}, {Schlickeiser}, {Schwanke}, {Sol}, {Spangler}, {Spanier},
  {Steenkamp}, {Stegmann}, {Superina}, {Tavernet}, {Terrier}, {Th{\'e}oret},
  {Tluczykont}, {van Eldik}, {Vasileiadis}, {Venter}, {Vincent}, {V{\"o}lk},
  {Wagner}, \& {Ward}}]{HESS:crab}
{Aharonian}, F., {Akhperjanian}, A.~G., {Bazer-Bachi}, A.~R., {et~al.}
  2006{\natexlab{a}}, \aap, 457, 899

\bibitem[{{Aharonian} {et~al.}(2006{\natexlab{b}}){Aharonian}, {Akhperjanian},
  {Bazer-Bachi}, {Beilicke}, {Benbow}, {Berge}, {Bernl{\"o}hr}, {Boisson},
  {Bolz}, {Borrel}, {Braun}, {Breitling}, {Brown}, {B{\"u}hler},
  {B{\"u}sching}, {Carrigan}, {Chadwick}, {Chounet}, {Cornils}, {Costamante},
  {Degrange}, {Dickinson}, {Djannati-Ata{\"i}}, {O'C.~Drury}, {Dubus},
  {Egberts}, {Emmanoulopoulos}, {Epinat}, {Espigat}, {Feinstein}, {Ferrero},
  {Fontaine}, {Funk}, {Funk}, {Gallant}, {Giebels}, {Glicenstein}, {Goret},
  {Hadjichristidis}, {Hauser}, {Hauser}, {Heinzelmann}, {Henri}, {Hermann},
  {Hinton}, {Hofmann}, {Holleran}, {Horns}, {Jacholkowska}, {de Jager},
  {Kh{\'e}lifi}, {Komin}, {Konopelko}, {Latham}, {Le Gallou}, {Lemi{\`e}re},
  {Lemoine-Goumard}, {Lohse}, {Martin}, {Martineau-Huynh}, {Marcowith},
  {Masterson}, {McComb}, {de Naurois}, {Nedbal}, {Nolan}, {Noutsos}, {Orford},
  {Osborne}, {Ouchrif}, {Panter}, {Pelletier}, {Pita}, {P{\"u}hlhofer},
  {Punch}, {Raubenheimer}, {Raue}, {Rayner}, {Reimer}, {Reimer}, {Ripken},
  {Rob}, {Rolland}, {Rowell}, {Sahakian}, {Saug{\'e}}, {Schlenker},
  {Schlickeiser}, {Schwanke}, {Sol}, {Spangler}, {Spanier}, {Steenkamp},
  {Stegmann}, {Superina}, {Tavernet}, {Terrier}, {Th{\'e}oret}, {Tluczykont},
  {van Eldik}, {Vasileiadis}, {Venter}, {Vincent}, {V{\"o}lk}, {Wagner}, \&
  {Ward}}]{HESS:VelaX}
{Aharonian}, F., {Akhperjanian}, A.~G., {Bazer-Bachi}, A.~R., {et~al.}
  2006{\natexlab{b}}, \aap, 448, L43

\bibitem[{{Aharonian} {et~al.}(2006{\natexlab{c}}){Aharonian}, {Akhperjanian},
  {Bazer-Bachi}, {Beilicke}, {Benbow}, {Berge}, {Bernl{\"o}hr}, {Boisson},
  {Bolz}, {Borrel}, {Braun}, {Breitling}, {Brown}, {Chadwick}, {Chounet},
  {Cornils}, {Costamante}, {Degrange}, {Dickinson}, {Djannati-Ata{\"i}},
  {Drury}, {Dubus}, {Emmanoulopoulos}, {Espigat}, {Feinstein}, {Fontaine},
  {Fuchs}, {Funk}, {Gallant}, {Giebels}, {Gillessen}, {Glicenstein}, {Goret},
  {Hadjichristidis}, {Hauser}, {Hauser}, {Heinzelmann}, {Henri}, {Hermann},
  {Hinton}, {Hofmann}, {Holleran}, {Horns}, {Jacholkowska}, {de Jager},
  {Kh{\'e}lifi}, {Klages}, {Komin}, {Konopelko}, {Latham}, {Le Gallou},
  {Lemi{\`e}re}, {Lemoine-Goumard}, {Leroy}, {Lohse}, {Marcowith}, {Martin},
  {Martineau-Huynh}, {Masterson}, {McComb}, {de Naurois}, {Nolan}, {Noutsos},
  {Orford}, {Osborne}, {Ouchrif}, {Panter}, {Pelletier}, {Pita},
  {P{\"u}hlhofer}, {Punch}, {Raubenheimer}, {Raue}, {Raux}, {Rayner}, {Reimer},
  {Reimer}, {Ripken}, {Rob}, {Rolland}, {Rowell}, {Sahakian}, {Saug{\'e}},
  {Schlenker}, {Schlickeiser}, {Schuster}, {Schwanke}, {Siewert}, {Sol},
  {Spangler}, {Steenkamp}, {Stegmann}, {Tavernet}, {Terrier}, {Th{\'e}oret},
  {Tluczykont}, {van Eldik}, {Vasileiadis}, {Venter}, {Vincent}, {V{\"o}lk}, \&
  {Wagner}}]{HESS:gc_diffuse}
{Aharonian}, F., {Akhperjanian}, A.~G., {Bazer-Bachi}, A.~R., {et~al.}
  2006{\natexlab{c}}, \nat, 439, 695

\bibitem[{{Aharonian} {et~al.}(2006{\natexlab{d}}){Aharonian}, {Akhperjanian},
  {Bazer-Bachi}, {Beilicke}, {Benbow}, {Berge}, {Bernl{\"o}hr}, {Boisson},
  {Bolz}, {Borrel}, {Braun}, {Breitling}, {Brown}, {Chadwick}, {Chounet},
  {Cornils}, {Costamante}, {Degrange}, {Dickinson}, {Djannati-Ata{\"i}},
  {Drury}, {Dubus}, {Emmanoulopoulos}, {Espigat}, {Feinstein}, {Fontaine},
  {Fuchs}, {Funk}, {Gallant}, {Giebels}, {Gillessen}, {Glicenstein}, {Goret},
  {Hadjichristidis}, {Hauser}, {Heinzelmann}, {Henri}, {Hermann}, {Hinton},
  {Hofmann}, {Holleran}, {Horns}, {Jacholkowska}, {de Jager}, {Kh{\'e}lifi},
  {Komin}, {Konopelko}, {Latham}, {Le Gallou}, {Lemi{\`e}re},
  {Lemoine-Goumard}, {Leroy}, {Lohse}, {Martin}, {Martineau-Huynh},
  {Marcowith}, {Masterson}, {McComb}, {de Naurois}, {Nolan}, {Noutsos},
  {Orford}, {Osborne}, {Ouchrif}, {Panter}, {Pelletier}, {Pita},
  {P{\"u}hlhofer}, {Punch}, {Raubenheimer}, {Raue}, {Raux}, {Rayner}, {Reimer},
  {Reimer}, {Ripken}, {Rob}, {Rolland}, {Rowell}, {Sahakian}, {Saug{\'e}},
  {Schlenker}, {Schlickeiser}, {Schuster}, {Schwanke}, {Siewert}, {Sol},
  {Spangler}, {Steenkamp}, {Stegmann}, {Tavernet}, {Terrier}, {Th{\'e}oret},
  {Tluczykont}, {Vasileiadis}, {Venter}, {Vincent}, {V{\"o}lk}, \&
  {Wagner}}]{HESS:scanpaper2}
{Aharonian}, F., {Akhperjanian}, A.~G., {Bazer-Bachi}, A.~R., {et~al.}
  2006{\natexlab{d}}, \apj, 636, 777

\bibitem[{{Aharonian} {et~al.}(2006{\natexlab{e}}){Aharonian}, {Akhperjanian},
  {Bazer-Bachi}, {Beilicke}, {Benbow}, {Berge}, {Bernl{\"o}hr}, {Boisson},
  {Bolz}, {Borrel}, {Braun}, {Breitling}, {Brown}, {Chadwick}, {Chounet},
  {Cornils}, {Costamante}, {Degrange}, {Dickinson}, {Djannati-Ata{\"i}},
  {O'C.~Drury}, {Dubus}, {Emmanoulopoulos}, {Espigat}, {Feinstein}, {Fontaine},
  {Fuchs}, {Funk}, {Gallant}, {Giebels}, {Glicenstein}, {Goret},
  {Hadjichristidis}, {Hauser}, {Hauser}, {Heinzelmann}, {Henri}, {Hermann},
  {Hinton}, {Hofmann}, {Holleran}, {Horns}, {Jacholkowska}, {de Jager},
  {Kh{\'e}lifi}, {Klages}, {Komin}, {Konopelko}, {Latham}, {Le Gallou},
  {Lemi{\`e}re}, {Lemoine-Goumard}, {Lohse}, {Martin}, {Martineau-Huynh},
  {Marcowith}, {Masterson}, {McComb}, {de Naurois}, {Nedbal}, {Nolan},
  {Noutsos}, {Orford}, {Osborne}, {Ouchrif}, {Panter}, {Pelletier}, {Pita},
  {P{\"u}hlhofer}, {Punch}, {Raubenheimer}, {Raue}, {Rayner}, {Reimer},
  {Reimer}, {Ripken}, {Rob}, {Rolland}, {Rowell}, {Sahakian}, {Saug{\'e}},
  {Schlenker}, {Schlickeiser}, {Schuster}, {Schwanke}, {Siewert}, {Sol},
  {Spangler}, {Steenkamp}, {Stegmann}, {Superina}, {Tavernet}, {Terrier},
  {Th{\'e}oret}, {Tluczykont}, {van Eldik}, {Vasileiadis}, {Venter}, {Vincent},
  {V{\"o}lk}, \& {Wagner}}]{HESS:RXJ1713}
{Aharonian}, F., {Akhperjanian}, A.~G., {Bazer-Bachi}, A.~R., {et~al.}
  2006{\natexlab{e}}, \aap, 449, 223

\bibitem[{{Aharonian} {et~al.}(2006{\natexlab{f}}){Aharonian}, {Akhperjanian},
  {Bazer-Bachi}, {Beilicke}, {Benbow}, {Berge}, {Bernl{\"o}hr}, {Boisson},
  {Bolz}, {Borrel}, {Braun}, {Brown}, {B{\"u}hler}, {B{\"u}sching}, {Carrigan},
  {Chadwick}, {Chounet}, {Cornils}, {Costamante}, {Degrange}, {Dickinson},
  {Djannati-Ata{\"i}}, {O'C.~Drury}, {Dubus}, {Egberts}, {Emmanoulopoulos},
  {Espigat}, {Feinstein}, {Ferrero}, {Fiasson}, {Fontaine}, {Funk}, {Funk},
  {F{\"u}{\ss}ling}, {Gallant}, {Giebels}, {Glicenstein}, {Goret},
  {Hadjichristidis}, {Hauser}, {Hauser}, {Heinzelmann}, {Henri}, {Hermann},
  {Hinton}, {Hoffmann}, {Hofmann}, {Holleran}, {Horns}, {Jacholkowska}, {de
  Jager}, {Kendziorra}, {Kh{\'e}lifi}, {Komin}, {Konopelko}, {Kosack},
  {Latham}, {Le Gallou}, {Lemi{\`e}re}, {Lemoine-Goumard}, {Lohse}, {Martin},
  {Martineau-Huynh}, {Marcowith}, {Masterson}, {Maurin}, {McComb}, {de
  Naurois}, {Nedbal}, {Nolan}, {Noutsos}, {Orford}, {Osborne}, {Ouchrif},
  {Panter}, {Pelletier}, {Pita}, {P{\"u}hlhofer}, {Punch}, {Raubenheimer},
  {Raue}, {Rayner}, {Reimer}, {Reimer}, {Ripken}, {Rob}, {Rolland}, {Rowell},
  {Sahakian}, {Santangelo}, {Saug{\'e}}, {Schlenker}, {Schlickeiser},
  {Schr{\"o}der}, {Schwanke}, {Schwarzburg}, {Shalchi}, {Sol}, {Spangler},
  {Spanier}, {Steenkamp}, {Stegmann}, {Superina}, {Tavernet}, {Terrier},
  {Th{\'e}oret}, {Tluczykont}, {van Eldik}, {Vasileiadis}, {Venter}, {Vincent},
  {V{\"o}lk}, {Wagner}, \& {Ward}}]{HESS:Kookaburra}
{Aharonian}, F., {Akhperjanian}, A.~G., {Bazer-Bachi}, A.~R., {et~al.}
  2006{\natexlab{f}}, \aap, 456, 245

\bibitem[{{Aharonian} {et~al.}(2005{\natexlab{d}}){Aharonian}, {Akhperjanian},
  {Bazer-Bachi}, {Beilicke}, {Benbow}, {Berge}, {Bernl{\"o}hr}, {Boisson},
  {Bolz}, {Borrel}, {Braun}, {Breitling}, {Brown}, {Chadwick}, {Chounet},
  {Cornils}, {Costamante}, {Degrange}, {Dickinson}, {Djannati-Ata{\"i}},
  {O'C.~Drury}, {Dubus}, {Emmanoulopoulos}, {Espigat}, {Feinstein}, {Fontaine},
  {Fuchs}, {Funk}, {Gallant}, {Giebels}, {Gillessen}, {Glicenstein}, {Goret},
  {Hadjichristidis}, {Hauser}, {Heinzelmann}, {Henri}, {Hermann}, {Hinton},
  {Hofmann}, {Holleran}, {Horns}, {Jacholkowska}, {de Jager}, {Kh{\'e}lifi},
  {Komin}, {Konopelko}, {Latham}, {Le Gallou}, {Lemi{\`e}re},
  {Lemoine-Goumard}, {Leroy}, {Lohse}, {Martin}, {Martineau-Huynh},
  {Marcowith}, {Masterson}, {McComb}, {de Naurois}, {Nolan}, {Noutsos},
  {Orford}, {Osborne}, {Ouchrif}, {Panter}, {Pelletier}, {Pita},
  {P{\"u}hlhofer}, {Punch}, {Raubenheimer}, {Raue}, {Raux}, {Rayner}, {Reimer},
  {Reimer}, {Ripken}, {Rob}, {Rolland}, {Rowell}, {Sahakian}, {Saug{\'e}},
  {Schlenker}, {Schlickeiser}, {Schuster}, {Schwanke}, {Siewert}, {Sol},
  {Spangler}, {Steenkamp}, {Stegmann}, {Tavernet}, {Terrier}, {Th{\'e}oret},
  {Tluczykont}, {Vasileiadis}, {Venter}, {Vincent}, {V{\"o}lk}, \&
  {Wagner}}]{HESS:J1825}
{Aharonian}, F.~A., {Akhperjanian}, A.~G., {Bazer-Bachi}, A.~R., {et~al.}
  2005{\natexlab{d}}, \aap, 442, L25

\bibitem[{{Aharonian} \& {Atoyan}(1996)}]{aharonian96:_emiss_of_pi_decay_gamma}
{Aharonian}, F.~A. \& {Atoyan}, A.~M. 1996, \aap, 309, 917

\bibitem[{{Aharonian} {et~al.}(1994){Aharonian}, {Drury}, \&
  {Voelk}}]{aharonian94:_gev_tev_gamma_ray_emiss}
{Aharonian}, F.~A., {Drury}, L.~O., \& {Voelk}, H.~J. 1994, \aap, 285, 645

\bibitem[{{Bamba} {et~al.}(2000){Bamba}, {Yokogawa}, {Sakano}, \&
  {Koyama}}]{bamba00:_g359}
{Bamba}, A., {Yokogawa}, J., {Sakano}, M., \& {Koyama}, K. 2000, \pasj, 52, 259

\bibitem[{{Berge} {et~al.}(2007){Berge}, {Funk}, \& {Hinton}}]{HESS:background}
{Berge}, D., {Funk}, S., \& {Hinton}, J. 2007, \aap, 466, 1219

\bibitem[{{Bernl{\"o}hr} {et~al.}(2003){Bernl{\"o}hr}, {Carrol}, {Cornils},
  {Elfahem}, {Espigat}, {Gillessen}, {Heinzelmann}, {Hermann}, {Hofmann},
  {Horns}, {Jung}, {Kankanyan}, {Katona}, {Khelifi}, {Krawczynski}, {Panter},
  {Punch}, {Rayner}, {Rowell}, {Tluczykont}, \& {van Staa}}]{HESS:optics}
{Bernl{\"o}hr}, K., {Carrol}, O., {Cornils}, R., {et~al.} 2003, Astroparticle
  Physics, 20, 111

\bibitem[{{Bird} {et~al.}(2007){Bird}, {Malizia}, {Bazzano}, {Barlow},
  {Bassani}, {Hill}, {B{\'e}langer}, {Capitanio}, {Clark}, {Dean}, {Fiocchi},
  {G{\"o}tz}, {Lebrun}, {Molina}, {Produit}, {Renaud}, {Sguera}, {Stephen},
  {Terrier}, {Ubertini}, {Walter}, {Winkler}, \& {Zurita}}]{INTEGRAL:3IBIS}
{Bird}, A.~J., {Malizia}, A., {Bazzano}, A., {et~al.} 2007, \apjs, 170, 175

\bibitem[{{Bitran} {et~al.}(1997){Bitran}, {Alvarez}, {Bronfman}, {May}, \&
  {Thaddeus}}]{bitran97:_large_scale_co_survey}
{Bitran}, M., {Alvarez}, H., {Bronfman}, L., {May}, J., \& {Thaddeus}, P. 1997,
  \aaps, 125, 99

\bibitem[{{Carrigan} {et~al.}(2007){Carrigan}, {Hinton}, {Hofmann}, {Kosack},
  {Lohse}, {Reimer}, \& {for the H.~E.~S.~S.~Collaboration}}]{carrigan07}
{Carrigan}, S., {Hinton}, J.~A., {Hofmann}, W., {et~al.} 2007, ArXiv e-prints,
  709

\bibitem[{{Dame} {et~al.}(2001){Dame}, {Hartmann}, \&
  {Thaddeus}}]{dame01:_milky_way_in_molec_cloud}
{Dame}, T.~M., {Hartmann}, D., \& {Thaddeus}, P. 2001, \apj, 547, 792

\bibitem[{{Daum} {et~al.}(1997){Daum}, {Hermann}, {Hess}, {Hofmann},
  {Lampeitl}, {P{\"u}hlhofer}, {Aharonian}, {Akhperjanian}, {Barrio},
  {Beglarian}, {Bernl{\"o}hr}, {Beteta}, {Bradbury}, {Contreras}, {Cortina},
  {Deckers}, {Feigl}, {Fernandez}, {Fonseca}, {Frass}, {Funk}, {Gonzalez},
  {Heinzelmann}, {Hemberger}, {Heusler}, {Holl}, {Horns}, {Kankanyan},
  {Kirstein}, {K{\"o}hler}, {Konopelko}, {Kranich}, {Krawczynski}, {Kornmayer},
  {Lindner}, {Lorenz}, {Magnussen}, {Meyer}, {Mirzoyan}, {M{\"o}ller},
  {Moralejo}, {Padilla}, {Panter}, {Petry}, {Plaga}, {Prahl}, {Prosch},
  {Rauterberg}, {Rhode}, {R{\"o}hring}, {Sahakian}, {Samorski}, {Sanchez},
  {Schmele}, {Stamm}, {Ulrich}, {V{\"o}lk}, {Westerhoff}, {Wiebel-Sooth},
  {Wiedner}, {Willmer}, \& {Wirth}}]{HEGRA:acts}
{Daum}, A., {Hermann}, G., {Hess}, M., {et~al.} 1997, Astroparticle Physics, 8,
  1

\bibitem[{{de Naurois}(2005)}]{HESS:Model2D}
{de Naurois}, M. 2005, in proceedings of the conference \emph{Towards a Network
  of Atmospheric Cherenkov Detectors VII}, Palaiseau, France, astro-ph/0607247

\bibitem[{{Downes} {et~al.}(1979){Downes}, {Goss}, {Schwarz}, \&
  {Wouterloot}}]{downes79:radio_gc}
{Downes}, D., {Goss}, W.~M., {Schwarz}, U.~J., \& {Wouterloot}, J.~G.~A. 1979,
  \aaps, 35, 1

\bibitem[{{Elitzur}(1976)}]{elitzur76:_inver_of_oh_mhz_line}
{Elitzur}, M. 1976, \apj, 203, 124

\bibitem[{{Funk} {et~al.}(2007){Funk}, {Reimer}, {Torres}, \&
  {Hinton}}]{funk07:_gev_tev_connec_galac}
{Funk}, S., {Reimer}, O., {Torres}, D.~F., \& {Hinton}, J.~A. 2007, ArXiv
  e-prints, 710

\bibitem[{{Gabici} \& {Aharonian}(2007)}]{gabici07}
{Gabici}, S. \& {Aharonian}, F.~A. 2007, \apjl, 665, L131

\bibitem[{{Gaensler} {et~al.}(2004){Gaensler}, {van der Swaluw}, {Camilo},
  {Kaspi}, {Baganoff}, {Yusef-Zadeh}, \&
  {Manchester}}]{gaensler04:_mouse_soared}
{Gaensler}, B.~M., {van der Swaluw}, E., {Camilo}, F., {et~al.} 2004, \apj,
  616, 383

\bibitem[{{Green}(2004)}]{green04:SNRs}
{Green}, D.~A. 2004, Bulletin of the Astronomical Society of India, 32, 335

\bibitem[{{Hartman} {et~al.}(1999){Hartman}, {Bertsch}, {Bloom}, {Chen},
  {Deines-Jones}, {Esposito}, {Fichtel}, {Friedlander}, {Hunter}, {McDonald},
  {Sreekumar}, {Thompson}, {Jones}, {Lin}, {Michelson}, {Nolan}, {Tompkins},
  {Kanbach}, {Mayer-Hasselwander}, {M{\"u}cke}, {Pohl}, {Reimer}, {Kniffen},
  {Schneid}, {von Montigny}, {Mukherjee}, \& {Dingus}}]{EGRET:3EG}
{Hartman}, R.~C., {Bertsch}, D.~L., {Bloom}, S.~D., {et~al.} 1999, \apjs, 123,
  79

\bibitem[{{Hillas}(1985)}]{hillas85}
{Hillas}, A.~M. 1985, in International Cosmic Ray Conference, Vol.~3,
  International Cosmic Ray Conference, ed. F.~C. {Jones}, 445--448

\bibitem[{{Hillas}(1996)}]{hillas96:technique}
{Hillas}, A.~M. 1996, Space Science Reviews, 75, 17

\bibitem[{{Hinton} \& {Aharonian}(2007)}]{hinton07}
{Hinton}, J.~A. \& {Aharonian}, F.~A. 2007, \apj, 657, 302

\bibitem[{{Hinton} {et~al.}(2007){Hinton}, {Funk}, {Carrigan}, {Gallant}, {de
  Jager}, {Kosack}, {Lemi{\`e}re}, \& {P{\"u}hlhofer}}]{hinton07:hessj1718_XMM}
{Hinton}, J.~A., {Funk}, S., {Carrigan}, S., {et~al.} 2007, ArXiv e-prints, 710

\bibitem[{{in~'t~Zand} {et~al.}(2007){in~'t~Zand}, {Jonker}, \&
  {Markwardt}}]{zand07:1744_299}
{in~'t~Zand}, J.~J.~M., {Jonker}, P.~G., \& {Markwardt}, C.~B. 2007, \aap, 465,
  953

\bibitem[{{Kosack} {et~al.}(2004){Kosack}, {Badran}, {Bond}, {Boyle},
  {Bradbury}, {Buckley}, {Carter-Lewis}, {Celik}, {Connaughton}, {Cui},
  {Daniel}, {D'Vali}, {de la Calle Perez}, {Duke}, {Falcone}, {Fegan}, {Fegan},
  {Finley}, {Fortson}, {Gaidos}, {Gammell}, {Gibbs}, {Gillanders}, {Grube},
  {Gutierrez}, {Hall}, {Hall}, {Hanna}, {Hillas}, {Holder}, {Horan}, {Jarvis},
  {Jordan}, {Kenny}, {Kertzman}, {Kieda}, {Kildea}, {Knapp}, {Krawczynski},
  {Krennrich}, {Lang}, {Le Bohec}, {Linton}, {Lloyd-Evans}, {Milovanovic},
  {McEnery}, {Moriarty}, {Muller}, {Nagai}, {Nolan}, {Ong}, {Pallassini},
  {Petry}, {Power-Mooney}, {Quinn}, {Quinn}, {Ragan}, {Rebillot}, {Reynolds},
  {Rose}, {Schroedter}, {Sembroski}, {Swordy}, {Syson}, {Vassiliev}, {Wakely},
  {Walker}, {Weekes}, \& {Zweerink}}]{kosack04}
{Kosack}, K., {Badran}, H.~M., {Bond}, I.~H., {et~al.} 2004, \apjl, 608, L97

\bibitem[{{LaRosa} {et~al.}(2000){LaRosa}, {Kassim}, {Lazio}, \&
  {Hyman}}]{larosa00:_wide_field_centim_vla_image}
{LaRosa}, T.~N., {Kassim}, N.~E., {Lazio}, T.~J.~W., \& {Hyman}, S.~D. 2000,
  \aj, 119, 207

\bibitem[{{Li} \& {Ma}(1983)}]{li_ma83}
{Li}, T.-P. \& {Ma}, Y.-Q. 1983, \apj, 272, 317

\bibitem[{{Liu} {et~al.}(2006){Liu}, {van Paradijs}, \& {van den
  Heuvel}}]{liu06:HMXB}
{Liu}, Q.~Z., {van Paradijs}, J., \& {van den Heuvel}, E.~P.~J. 2006, \aap,
  455, 1165

\bibitem[{{Manchester} {et~al.}(2005){Manchester}, {Hobbs}, {Teoh}, \&
  {Hobbs}}]{ATNF}
{Manchester}, R.~N., {Hobbs}, G.~B., {Teoh}, A., \& {Hobbs}, M. 2005, \aj, 129,
  1993

\bibitem[{{Piron} {et~al.}(2001){Piron}, {Djannati-Atai}, {Punch}, {Tavernet},
  {Barrau}, {Bazer-Bachi}, {Chounet}, {Debiais}, {Degrange}, {Dezalay},
  {Espigat}, {Fabre}, {Fleury}, {Fontaine}, {Goret}, {Gouiffes}, {Khelifi},
  {Malet}, {Masterson}, {Mohanty}, {Nuss}, {Renault}, {Rivoal}, {Rob}, \&
  {Vorobiov}}]{piron01:CAT_ForwardFolding_Mrk421}
{Piron}, F., {Djannati-Atai}, A., {Punch}, M., {et~al.} 2001, \aap, 374, 895

\bibitem[{{Robinson} {et~al.}(1996){Robinson}, {Yusef-Zadeh}, \&
  {Roberts}}]{robinson96:_mhz_oh_maser_emiss_snr_g359}
{Robinson}, B., {Yusef-Zadeh}, F., \& {Roberts}, D. 1996, in Bulletin of the
  American Astronomical Society, Vol.~28, 948--+

\bibitem[{{Skinner} {et~al.}(1990){Skinner}, {Foster}, {Willmore}, \&
  {Eyles}}]{skinner90}
{Skinner}, G.~K., {Foster}, A.~J., {Willmore}, A.~P., \& {Eyles}, C.~J. 1990,
  \mnras, 243, 72

\bibitem[{{Snowden} {et~al.}(2004){Snowden}, {Collier}, \&
  {Kuntz}}]{snowden04:ESAS}
{Snowden}, S.~L., {Collier}, M.~R., \& {Kuntz}, K.~D. 2004, \apj, 610, 1182

\bibitem[{{Tanaka} {et~al.}(1994){Tanaka}, {Inoue}, \& {Holt}}]{ASCA}
{Tanaka}, Y., {Inoue}, H., \& {Holt}, S.~S. 1994, \pasj, 46, L37

\bibitem[{{Torres} {et~al.}(2001){Torres}, {Romero}, {Combi}, {Benaglia},
  {Andernach}, \& {Punsly}}]{torres01:_variab_analy_of_low_latit}
{Torres}, D.~F., {Romero}, G.~E., {Combi}, J.~A., {et~al.} 2001, \aap, 370, 468

\bibitem[{{Torres} {et~al.}(2003){Torres}, {Romero}, {Dame}, {Combi}, \&
  {Butt}}]{torres03:_super_remnan_and_gamma_ray_sourc}
{Torres}, D.~F., {Romero}, G.~E., {Dame}, T.~M., {Combi}, J.~A., \& {Butt},
  Y.~M. 2003, \physrep, 382, 303

\bibitem[{{Tsuchiya} {et~al.}(2004){Tsuchiya}, {Enomoto}, {Ksenofontov},
  {Mori}, {Naito}, {Asahara}, {Bicknell}, {Clay}, {Doi}, {Edwards}, {Gunji},
  {Hara}, {Hara}, {Hattori}, {Hayashi}, {Itoh}, {Kabuki}, {Kajino}, {Katagiri},
  {Kawachi}, {Kifune}, {Kubo}, {Kurihara}, {Kurosaka}, {Kushida}, {Matsubara},
  {Miyashita}, {Mizumoto}, {Moro}, {Muraishi}, {Muraki}, {Nakase}, {Nishida},
  {Nishijima}, {Ohishi}, {Okumura}, {Patterson}, {Protheroe}, {Sakamoto},
  {Sakurazawa}, {Swaby}, {Tanimori}, {Tanimura}, {Thornton}, {Tokanai},
  {Uchida}, {Watanabe}, {Yamaoka}, {Yanagita}, {Yoshida}, \&
  {Yoshikoshi}}]{tsuchiya04:_detec_sub_tev_gamma_rays}
{Tsuchiya}, K., {Enomoto}, R., {Ksenofontov}, L.~T., {et~al.} 2004, \apjl, 606,
  L115

\bibitem[{{Uchida} {et~al.}(1992{\natexlab{a}}){Uchida}, {Morris}, \&
  {Yusef-Zadeh}}]{uchida92:_h_i_absor_line_study}
{Uchida}, K., {Morris}, M., \& {Yusef-Zadeh}, F. 1992{\natexlab{a}}, \aj, 104,
  1533

\bibitem[{{Uchida} {et~al.}(1992{\natexlab{b}}){Uchida}, {Morris}, {Bally},
  {Pound}, \& {Yusef-Zadeh}}]{uchida92:_dense_molec_ring_surroun_nonth}
{Uchida}, K.~I., {Morris}, M., {Bally}, J., {Pound}, M., \& {Yusef-Zadeh}, F.
  1992{\natexlab{b}}, \apj, 398, 128

\bibitem[{{van der Hucht}(2001)}]{hucht01:_catal_galac_wolf_rayet}
{van der Hucht}, K.~A. 2001, VizieR Online Data Catalog, 3215, 0

\bibitem[{{Voges} {et~al.}(2000){Voges}, {Aschenbach}, {Boller}, {Brauninger},
  {Briel}, {Burkert}, {Dennerl}, {Englhauser}, {Gruber}, {Haberl}, {Hartner},
  {Hasinger}, {Pfeffermann}, {Pietsch}, {Predehl}, {Schmitt}, {Trumper}, \&
  {Zimmermann}}]{ROSAT}
{Voges}, W., {Aschenbach}, B., {Boller}, T., {et~al.} 2000, \iaucirc, 7432, 3

\bibitem[{{Weekes}(1996)}]{weekes96:acts}
{Weekes}, T.~C. 1996, Space Science Reviews, 75, 1

\bibitem[{{Yusef-Zadeh} {et~al.}(1995){Yusef-Zadeh}, {Uchida}, \&
  {Roberts}}]{yusef-zadeh95:_shock_excit_oh_maser_emiss}
{Yusef-Zadeh}, F., {Uchida}, K.~I., \& {Roberts}, D. 1995, Science, 270, 1801

\end{thebibliography}

\end{document}